\documentclass{bmvc2k}

\usepackage[misc]{ifsym}

\title{TransResNet: Integrating the Strengths of ViTs and CNNs for High Resolution Medical Image Segmentation via Feature Grafting}

\addauthor{Muhammad Hamza Sharif}{muhammad.sharif@mbzuai.ac.ae}{1}
\addauthor{Dmitry Demidov}{dmitry.demidov@mbzuai.ac.ae}{1}
\addauthor{Asif Hanif}{asif.hanif@mbzuai.ac.ae}{1}
\addauthor{Mohammad Yaqub}{mohammad.yaqub@mbzuai.ac.ae}{1}
\addauthor{Min Xu \Letter}{xumin100@gmail.com}{1}

\addinstitution{
Mohamed Bin Zayed University\\ of Artificial Intelligence, UAE
}
\runninghead{SHARIF ET AL}{TransResNet}

\def\etal{\emph{et al}\bmvaOneDot}

\usepackage{times}
\usepackage{epsfig}
\usepackage{amsmath}
\usepackage{amssymb}
\usepackage{pifont}
\usepackage{makecell}
\usepackage[toc,page]{appendix}
\usepackage{color, colortbl}

\usepackage{makecell}
\usepackage[rightcaption]{sidecap} % <-- added
\usepackage{bm}
\usepackage{physics}
\usepackage{pbox}

\definecolor{lightsalmon}{rgb}{1.0, 0.63, 0.48}
\definecolor{navajowhite}{rgb}{1.0, 0.87, 0.68}
\definecolor{mediumspringbud}{rgb}{0.79, 0.86, 0.54}
\definecolor{gray}{rgb}{0.75, 0.75, 0.75}
\newcommand{\X}{\mathrm{X}}
\newcommand{\Z}{\mathrm{Z}}
\newcommand{\W}{\mathrm{W}}

\newcommand{\Q}{\mathrm{Q}}
\newcommand{\K}{\mathrm{K}}
\newcommand{\V}{\mathrm{V}}

\newcommand{\Rf}{\mathrm{R}}
\newcommand{\Sf}{\mathrm{S}}

\newcommand{\cmark}{\ding{51}}%
\newcommand{\xmark}{\ding{55}}%
\usepackage{textcomp}
\usepackage{bm}
\usepackage{textcomp}
\def\R{\mathbb{R}}

\newcommand{\Y}{\mathrm{Y}}

\begin{document}

\maketitle

\begin{abstract}
High-resolution images are preferable in medical imaging domain as they significantly improve the diagnostic capability of the underlying method. In particular, high resolution helps substantially in improving automatic image segmentation. However, most of the existing deep learning-based techniques for medical image segmentation are optimized for input images having small spatial dimensions and perform poorly on high-resolution images. To address this shortcoming, we propose a parallel-in-branch architecture called TransResNet, which incorporates Transformer and CNN in a parallel manner to extract features from multi-resolution images independently. In TransResNet, we introduce Cross Grafting Module (CGM), which generates the \textit{grafted features}, enriched in both global semantic and low-level spatial details, by combining the feature maps from Transformer and CNN branches through fusion and self-attention mechanism. Moreover, we use these grafted features in the decoding process, increasing the information flow for better prediction of the segmentation mask. Extensive experiments on ten datasets demonstrate that TransResNet achieves either state-of-the-art or competitive results on several segmentation tasks, including skin lesion, retinal vessel, and polyp segmentation. The source code and pre-trained models are available at \url{https://github.com/Sharifmhamza/TransResNet}. \vspace{-1em}
\end{abstract}

%-------------------------------------------------------------------------
\section{Introduction}\label{sec:intro}\vspace{-1.0em}
Segmentation is a fundamental problem in the domain of computer vision with numerous practical applications, particularly in biomedical imaging analysis. Medical segmented images can be used in a wide range of applications, such as disease localization \cite{sharma2010automated}, tissue volume estimation \cite{wang1998quantification}, and studying anatomical structure \cite{pham2000survey}. Accurate and precise segmentation of medical images is a challenging task due to the nature of the complexity of 2D and 3D structures. Recent studies have demonstrated deep learning-based techniques as a powerful building block to accomplish this task accurately \cite{dong2019neural, shamshad2022transformers, asgari2021deep}.   

Resolution of an image in medical diagnosis plays an important role. In general, high-resolution images improve the results of a diagnostic method to determine the presence of certain diseases. A high-resolution image contains rich semantic information and it provides better chances for the extraction of useful information for a downstream task e.g. segmentation \cite{super}. Many deep learning-based approaches have been proposed to perform automatic medical image segmentation, such as segmenting organs \cite{gibson2018organ}, lesions \cite{wang2021boundary}, and tumors \cite{hatamizadeh2022swin}. However, these existing techniques are primarily designed to segment low-resolution (small spatial dimension) images, and they do not provide favorable results on high-resolution images due to discrepancies between sampling depth and receptive field size. With rapid technological revolution, medical image-capturing devices have undergone extensive modifications and advancements in recent years. Compared to the previous appurtenances, these modern devices are capable of capturing images with higher resolution. This requires the demand for deep learning-based segmentation framework that can process high-resolution medical images efficiently and performs favorably.

Encoder-decoder based convolutional neural network (CNN) architectures have achieved unprecedented performance in medical image segmentation for low resolution input images \cite{unet++, attentionunet, recurrentunet}. Despite their impressive success, these approaches are still facing challenges to capture global context details due to narrow and fixed receptive field. Similarly, vision transformers (ViTs) \cite{dosovitskiy2020image, swin}, which are efficient for modeling long-range dependencies and highly parallelizable, are computationally prohibitive and have to down-sample the image before processing. Because of the shortcomings of these architectures regarding high-resolution images, the substantial solution is to design a single architecture that collectively captures rich local and global information without increasing the computational complexity associated with high-resolution images and gives accurate segmentation results. 

\begin{figure*}
\begin{center}
\includegraphics[width=1\textwidth]{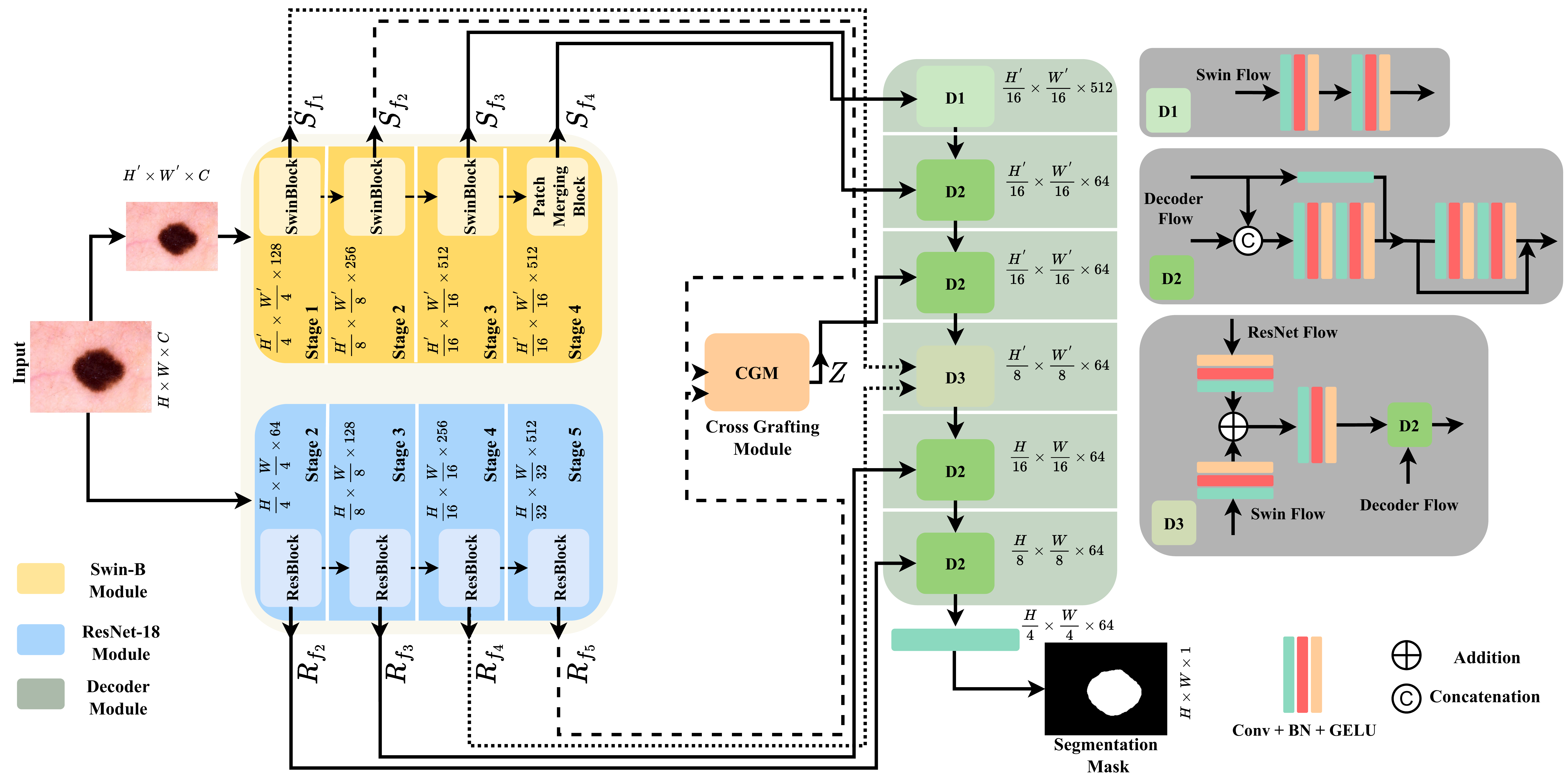}
\end{center}
\vspace{-2em}
\caption{\footnotesize{An overview of the architecture of TransResNet for high-resolution medical image segmentation. Our TransResNet uses the parallel branches from Swin-transformer and Resnet-18 backbones as encoders. The core module of our architecture is Cross Grafting Module (CGM), explained briefly in the Fig. \ref{fig:grafting}. The decoder module aggregates the flow of feature input maps from swin block, CGM block, and resnet block. D1, D2, and D3 are subblocks of the decoder with their structure on the right side.} } \vspace{-2.0em}
\label{fig:arch}
\end{figure*}

Inspired by deep learning-based high saliency object detection methods \cite{zeng2019towards, tang2021disentangled, xie2022pyramid} for natural images, we propose an architecture for high-resolution segmentation of medical images named \textbf{TransResNet}, as shown in Fig. \ref{fig:arch}. In this paper, we use two encoder modules, one is CCN-based for extracting local feature details, the other is transformer-based for grasping global information. We introduce a \emph{Cross Grafting Module \textbf{(CGM)}} to combine the features maps with similar spatial size from both encoder branches. CGM  generates \textit{grafted features} which are enriched in both local and global semantic cues. We use these grafted features in the decoding process for the prediction of segmentation masks. In summary, our main contributions are as follows:\vspace{-0.5em}
\begin{itemize}
\item We propose a framework named \textbf{TransResNet} for efficient segmentation of high-resolution medical images by using two encoder backbones.\vspace{-0.5em}

\item We introduce \emph{Cross-Grafting Module (CGM)}, to combine the low-level spatial features (from the CNN branch) and high level semantic information (from the Transformer branch) through fusion and self-attention mechanism.\vspace{-0.5em}

     \item We have performed extensive experiments on 
    ten datasets for three medical image segmentation tasks. Our experimental results demonstrate that the proposed approach outperforms state-of-the-art (SOTA) methods on high-resolution medical imaging datasets and competes comparably  on datasets containing mixture of low and high resolution images. \vspace{-0.5em}
\end{itemize}

\section{Related Work}\label{sec:related-work}\vspace{-1.0em}
Segmentation is not a trivial problem, especially in the biomedical imaging domain. A lot of studies have been proposed in the past that focus on low-level image information to predict the segmentation mask. The inefficiency of these methods in capturing rich semantics makes their performance inconsistent in complex settings. In this regard, we discuss a few works on biomedical image segmentation.\newline
\noindent \textbf{Medical Image Segmentation with CNN-based Architectures:} Extensive studies have been proposed predominantly with various CNN-based architectures for medical image segmentation tasks. Some studies have designed the model architecture in an encoder-decoder style, for example, U-net \cite{unet} and its variants \cite{unet++, attentionunet, recurrentunet}, while other works have integrated CNN extracted features with a self-attention mechanism in the decoder module to boost the network performance for capturing global interaction \cite{pranet, sarnet}. Despite their excellent performance, these approaches are limited in their ability to capture global semantic information due to narrow receptive fields, as the kernel size of CNN-based techniques is typically fixed, making it more challenging to predict segmentation masks accurately. Our work proposes a method which allows the receptive field of the CNN backbone to capture rich semantic information from high-resolution input images.

%\setdf{content={\tiny{This figure is switched off (to reduce compilation time of latex document).}}}
\noindent \textbf{Medical Image Segmentation with Vision Transformers:} Transformers, which were 
  primarily developed for the natural language processing task \cite{vaswani2017attention}, have made marvelous achievements in the field of computer vision for downstream tasks \cite{dosovitskiy2020image, swin}. Despite transformer\textquotesingle s powerful global modeling capabilities, they are limited in explicitly capturing rich semantic details that are essential for biomedical imaging analysis \cite{chen2021transunet}. The first transformer-based medical image segmentation framework named TransUnet, proposed by Chen \etal \cite{chen2021transunet}, uses a transformer backbone in the U-Net style to extract global features in the encoder block and upsample these encoded features in the decoder block. In the TransFuse framework, Zhang \etal \cite{zhang2021transfuse} combines CNN and Transformer in a parallel manner to grasp the local and global information for a similar task. Despite their success, these methods are suitable for low-resolution images and tend to ignore local semantic information in high-resolution images. Thus, our proposed architecture aims to mitigate this issue by incorporating novel \emph{Cross Grafting Module (CGM)} which captures both global and rich semantic details in high-resolution medical images.\vspace{-1.0em}

\section{Methodology}
\label{sec:methodology}\vspace{-0.5em}
The concrete architecture of proposed network is shown in the Fig. \ref{fig:arch}. The proposed network follows the encoder-decoder design architecture consisting of two encoders and one decoder. The encoder includes the ResNet-18 \cite{resnet} and Swin-B \cite{swin} as backbones, while decoding phase comprises of three sub-stages. The feature maps from both encoders are grafted in a Cross Grafting Module (CGM), which emphasises on the salient regions, and make network learn precise pixel-level details. We will discuss each part in the following subsections.\vspace{-1.0em}

\subsection{Encoder Module}\label{subsection:encoder}\vspace{-0.5em}
As explained earlier, our proposed architecture is based upon two encoding streams: a CNN and a vision transformer (ViT). The main reason for using two encoding streams is to capture both local and global information, which makes the network learn salient features more accurately. The CNN-based encoder captures the low-level feature representations from high-resolution input images, while the ViT-based encoder is used to learn the global semantic information from low-resolution input images as demonstrated in Fig. \ref{fig:arch}. During the encoding phase, images $(I \in \R^{H \times W \times C})$ having different spatial dimensions are passed to both encoders:  $I_{R} \in \R^{1024 \times 1024 \times 3}$ and $I_{S} \in \R^{224 \times 224 \times 3}$ are passed through ResNet-18 and Swin-B encoders respectively. Using two encoder networks with multi-scale input size consumes massive amount of computational resources to generate feature maps. To handle this issue, we have discarded some layers from Resnet-18, and Swin-B network. With prior knowledge from the literature \cite{resnet}, we know that there are five feature maps generated by Resnet-18, denoted as $\{\Rf_{f_{i}} | i  \in  (1,2,3,4, 5)\}$. The first stage of Resnet-18 uses a large kernel size of $7\times7$ to extract feature maps. In our case, the input $(I_{R} \in \R^{1024 \times 1024 \times 3})$ passing through Resnet-18, is larger in size, thus resulting in huge demand of computational capacity to generate feature maps for this stage. We relinquish this stage while keeping the last four stages of Resnet-18, $\{ \Rf_{f_{i}} | i  \in  (2,3,4,5) \}$, which learn more complex features and become computationally less expensive due to gradual down-sampling of the feature maps at each stage, resulting in $\{ \Rf_{f_{i}} \in \mathbb{R}^{  \frac{H}{2^{i}} \times \frac{W}{2^{i}} \times ( 32 \times 2^{i-1})}\}_{i=2}^{5}$ . In a similar fashion, we also follow similar approach with the Swin-B transformer. As there are four stages in Swin-B, we drop the last stage after the patch merging block. We utilize the feature maps generated from the output of first three stages and patch merging block of the fourth stage of Swin-B, denoted as $\{\Sf_{f_{i}} | i   \in   (1,2,3,4)\}$, with their feature embedding dimension represented as $\{ \Sf_{f_{i}} \in \mathbb{R}^{\frac{56}{2^{i-1}} \times \frac{56}{2^{i-1}} \times ( 64 \times 2^{i} )} \}_{i=1}^{3}$, and $\Sf_{f_{4}} \in \mathbb{R}^{14 \times 14 \times 512}$. As the spatial dimensions of feature maps $\Rf_{f_{5}}~\text{with dimension}~(32 \times 32 \times 512)$, and $\Sf_{f_{2}}~\text{with dimension}~(28 \times 28 \times 256)$, are very close to each other, we select these features for grafting in the CGM.\vspace{-1em}
%\iffalse
\begin{figure*}
\begin{center}
\includegraphics[width=0.90\textwidth]{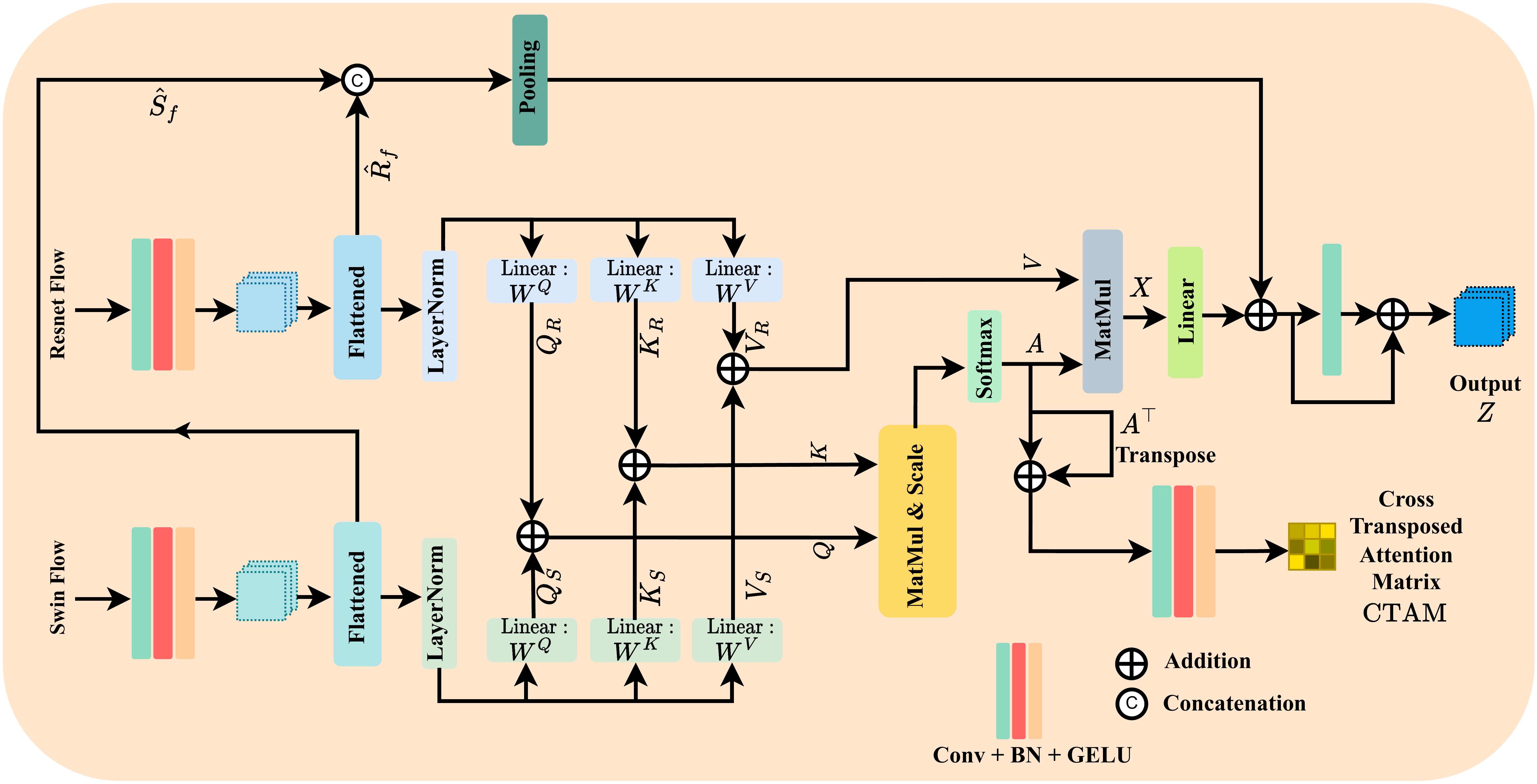}
\end{center}
\vspace{-2em}
\caption{\footnotesize{An overview of the architecture of the proposed Cross Grafting Module (CGM). The CGM module takes dual input i.e., the feature maps from Swin-transformer, and Resnet branches, and outputs the grafted features through fusion and self-attention mechanism. These grafted features are used in the decoding process. The module also generates a cross-transposed attention matrix (CTAM), which is used in the objective function.}}
\label{fig:grafting}\vspace{-2em}
\end{figure*}
%\fi

\subsection{Cross Grafting Module (CGM)}\label{subsection:CGM}\vspace{-0.5em}
We propose a new module, \emph{Cross Grafting Module (CGM)}, in such way that network effectively adapts the local and global semantic representations. To accomplish this task, we select the features maps $\Rf_{f_{5}} \in\R^{H^{'} \times W^{'} \times C^{'}}$, and $\Sf_{f_{2}} \in\R^{H^{''} \times W^{''} \times C^{''}}$ extracted from the Resnet-18 and Swin-B encoders respectively.  As transformers have shown excellence performance at modeling long range relationships \cite{dosovitskiy2020image}, $\Sf_{f_{2}}$ is responsible for providing global semantic detail, while $\Rf_{f_{5}}$ contributes towards local information due to CNN\textquotesingle s excellent low-level feature learning capabilities. But the major issue is the discrepancy between receptive field of feature maps, as $\Rf_{f_{5}}$ is down-sampled to match with receptive dimension of $\Sf_{f_{2}}$, thus produces the noisy output.

To alleviate this issue, we apply operator $\mathcal{A}$ (defined below) on the feature maps $(\Rf_{f_{5}} \in\R^{H^{'} \times W^{'} \times C^{'}}$, $\Sf_{f_{2}} \in\R^{H^{''} \times W^{''} \times C^{''}})$ and obtain $(\hat{\Rf}_{f} \in\R^{1 \times H^{'}W^{'}C^{'}}$, $\hat{\Sf}_{f} \in\R^{1 \times H^{''}W^{''}C^{''}})$ (see Eq. \ref{eq1}). Next, we apply layer normalization on $(\hat\Rf_f,\hat\Sf_f)$ and generate \textit{query} $(\Q)$, \textit{key} $(\K)$, and \textit{value} $(\V)$ projections from transformer and CNN branches separately (see Eq. \ref{eq2} and Eq. \ref{eq3}). By fusing (element-wise addition) the tensors in each tuple i.e. $(\Q_R,\Q_S), (\K_R,\K_S), (\V_R,\V_S)$, we achieve the resultant tensors with enriched local details, thus mitigating the effects of noise. For learning of global semantic information, we apply the self-attention (SA) mechanism that efficiently calculates the point-wise relationship between these resultant tensors. Grafted features $(\Z)$ i.e. the output of CGM, is used in the decoding process as shown in Fig. \ref{fig:arch}, but spatial dimension of SA output $(\X)$ has contradiction with the input dimension of decoder flow. To make it identical, we apply linear projection layer to SA output, reshape it to original size and feed it to convolution layer. During the whole process of grafting, the spatial dimension keeps on changing, motivating us to use shortcut connections. In order to enhance information flow and facilitate the training process, we use two skip connections before final output $(\Z)$ as shown in the Fig. \ref{fig:grafting}. Overall, the whole grafting process with intermediate steps is expressed as follows:\vspace{-1.0em}
\begin{equation}\label{eq1}
\begin{aligned}
\hat{\Rf}_{f} = \mathcal{A}(\Rf_{f_{5}})~~~&;~~~ 
\hat{\Sf}_{f} = \mathcal{A}(\Sf_{f_{2}}), \\
\hat{\Rf} = \mathit{LN}(\hat\Rf_f)~~~&;~~~ 
\hat{\Sf} = \mathit{LN}(\hat\Sf_f),
\end{aligned}
\end{equation}\vspace{-0.5em}
\begin{equation}\label{eq2}
\begin{aligned}
\Q_R &= \W^{Q}\hat{\Rf}~~~;~~~
\K_R = \W^{K}\hat{\Rf}~~~;~~~
\V_R = \W^{V}\hat{\Rf}, \\
\Q_S &= \W^{Q}\hat{\Sf}~~~~;~~~
\K_S = \W^{K}\hat{\Sf}~~~~;~~~
\V_S = \W^{V}\hat{\Sf},
\end{aligned}
\end{equation}\vspace{-0.5em}
\begin{equation}\label{eq3}
\Q = \Q_R + \Q_S~~;~~ 
\K = \K_R + \K_S~~;~~
\V = \V_R + \V_S,
\end{equation}\vspace{-1em}
\begin{equation}\label{eq4}
\begin{aligned}
\X &= \V \cdot \mathit{softmax} \left(\Q \cdot \K^{\top} / \alpha \right), \\
\Y &= \mathit{linear}(\X) + \mathit{pool}(\hat{\Rf}_{f} \oplus \hat{\Sf}_{f}), \\
\Z &= \Y + \mathit{conv}(\Y),
\end{aligned}
\end{equation}
\noindent where $\Rf_{f_{5}}$ and $\Sf_{f_{2}}$ are the input feature maps and $\mathcal{A} = \mathrm{Flatten}\leftarrow \mathrm{GELU} \leftarrow \mathrm{BN} \leftarrow \mathrm{Conv} (\cdot)$ is an operator that takes an input and performs convolution, batch normalization, GELU activation and flattening layer sequentially. Grafted feature $(\Z)$ is the output of CGM used in the decoding process. Here, $(\W^Q,\W^K,\W^V)$ are the weights of linear layers, $\alpha$ is a scaling parameter used to control the magnitude of dot product of $\Q$ and $\K$ tensors, and $(LN$, $\mathit{linear}$, $\mathit{pool})$ represent layer-normalization, linear and pooling layers respectively. $\oplus$ denotes the concatenation used to amalgamate the flatten output of transformer and resnet branches. Additionally, the dot product interaction of query and key projections generates the attention matrix $\textbf{A}=\mathrm{softmax} \left(\Q \cdot \K^{\top} / \alpha \right)$, which is transposed and added to the itself to form a cross transposed attention matrix which is defined by Eq. \ref{eq5} as follows:
\begin{equation}\label{eq5}
    \mathrm{\textbf{CTAM}} = \mathrm{GELU}(\mathrm{BN}(\mathrm{Conv}(\textbf{A} + \textbf{A}^{{\top}}))).
\end{equation}
$\mathrm{\textbf{CTAM}}$ matrix is used in the objective function (see Sec. \ref{subsection:objfunc}).\vspace{-1.0em}

\subsection{Decoder Module}\label{subsection:decoder}\vspace{-0.5em}
The flow of feature maps in the decoder block of our proposed architecture is illustrated in Fig.  \ref{fig:arch}. The decoder module first receives the flow of features from the Swin-B branch, followed by the feature grafting module, and finally the ResNet-18 branch in a staggered pattern. The decoder block is divided into chunks of three different sub-blocks, denoted as \textbf{D1}, \textbf{D2}, and \textbf{D3}, which aggregate the flow of features from these ResNet-18, Swin-B and CGM branches.\vspace{-1.0em}

\subsection{Objective Function}\label{subsection:objfunc}\vspace{-0.5em}
The entire network is trained end-to-end with the joint objective function, which includes the segmentation loss $L_{seg}$ for segmentation maps, attention loss $L_{att}$ for the cross-transposed attention matrix (\textbf{CTAM}) map , and auxiliary loss $L_{aux}$ for deep supervision to improve the gradient flow by supervising the ResNet-18 and Swin-B transformer branches, which is defined as follows: \vspace{-1em}
\begin{equation}\label{eq6}
\begin{aligned}
L_{seg} &= \frac{1}{2} \{ \phi_{bce}(\hat{M}_{pm}, M_{gt}) + \phi_{iou}(\hat{M}_{pm}, M_{gt}) \}, 
\end{aligned}
\end{equation}\vspace{-1.5em}
\begin{equation}\label{eq7}
\begin{aligned}
L_{att} &= \phi^w_{bce}(\mathrm{\textbf{{CTAM}}}_{map} , M_{gt_{map}}),
\end{aligned}
\end{equation}\vspace{-2em}
\begin{equation}\label{eq8}
\begin{aligned}
L_{aux} &= \frac{1}{2} \{ \phi_{bce}(\hat{M}_{R}, M_{gt}) + \phi_{iou}(\hat{M}_{R}, M_{gt}) \} + \frac{1}{2} \{ \phi_{bce}(\hat{M}_{S}, M_{gt}) + \phi_{iou}(\hat{M}_{S}, M_{gt}) \},
\end{aligned}
\end{equation}\vspace{-1.5em}
\begin{equation}\label{eq9}
\begin{aligned}
L_{total} &= L_{seg} + L_{att} + \lambda L_{aux},
\end{aligned}
\end{equation}
\noindent where $\phi_{bce}$, $\phi_{iou}$, and $\phi^w_{bce}$ denote the binary cross entropy function, intersection-over-union function, and weighted binary cross entropy function respectively.
Here, $M_{gt}$, $\hat{M}_{pm}$, $(\hat{M}_{R}$, $\hat{M}_{S})$, and $M_{gt_{map}}$ are ground truth mask, predicted segmentation mask, salient prediction maps extracted from the transformer, and resnet branches which are used in the grafting module, and attention matrix map generated from ground truth. Ground truth attention matrix map $M_{gt_{map}}$ is achieved by matching the shape of the ground truth mask $(M_{gt})$ to that of \textbf{CTAM} via down-sampling, flattening, and taking the self dot-product of flattened vector. The $\lambda$ is the weight parameter used to balance the auxiliary loss $L_{aux}$ calculated from two encoder branches.\vspace{-1.0em} 
\section{Experiments}\label{sec:experiments}\vspace{-0.5em}
We evaluate the effectiveness of our proposed model TransResNet on the three different segmentation tasks: \textbf{(a)} skin lesion segmentation (2 datasets),  \textbf{(b)} retinal vessel segmentation (3 datasets), and \textbf{(c)} polyp segmentation (5 datasets). More details on datasets, training settings, and additional quantitative and visual results have been presented in the following subsections. We have highlighted the best and the second-best highest scores in the result Sec. \ref{sec:quantitative-results} with different evaluation metrics.\vspace{-1.5em}
\subsection{Datasets}\label{subsec:datasets}\vspace{-0.5em}
\noindent \textbf{Skin Lesion Segmentation:} In order to segment skin lesions, we have used two publicly available benchmark datasets with the majority of high-resolution images (resolution $>$ 1K): ISIC-2016 \cite{isic2016}, and PH2 \cite{ph2}. The ISIC-2016 has the train-validation split of samples: 900/379, while the PH2 database includes 200 samples. As most of the SOTA methods use the same split, therefore, we also keep the same sample size for the fair evaluation of our model. To test the generalization ability and robustness of our method, we use PH2 dataset.\newline
\noindent \textbf{Retinal Vessel Segmentation:} The proposed method is evaluated using three publicly accessible retinal fundus imaging datasets: HRF \cite{hrf}, IOSTAR \cite{iostar}, and CHASE\_DB1 \cite{cashdb1}. The HRF database has a total of 45 samples, each image with a resolution of  $3504 \times 2306$, whereas IOSTAR contains 30, and CHASE\_DB1 has 28 image samples, respectively. For fair comparison, we also follow the same number of train-test split as mentioned in these papers \cite{m2u, fensemble, dunet, mdeep}.\newline
\noindent \textbf{Polyp Segmentation:} A total of five polyp segmentation benchmark datasets are used to evaluate the performance of TransResNet: Kvasir \cite{kvasir}, CVC-ClinicDB \cite{clinicdb}, CVC-ColonDB \cite{colondb}, Endoscene \cite{endoscene}, and ETIS \cite{etis}. To ensure fairness, we keep the same number of training and testing images as used in \cite{sarnet, pranet, zhang2021transfuse}, i.e. 1450 training images from Kvasir, and CVC-ClinicDB, while 798 testing images from all of the five datasets. These datasets are highly versatile benchmark datasets with mixture of low and high resolution images.\vspace{-1.5em}

\subsection{Implementation Details}\label{subsec:implementaion}\vspace{-0.5em}
We have implemented TransResNet using the Pytorch \cite{pytorch} framework and NVIDIA A100-SXM4 GPU with a maximum of 36GB of memory to accelerate the smooth training pipeline. All the input images have been resized to 1024 $\times$ 1024, and various data augmentations have been applied, including horizontal flip, vertical flip, rotation, and random brightness to increase the data diversity, volume, and avoid overfitting. The entire network is trained end-to-end with a Stochastic Gradient Optimizer (SGD) \cite{sgd} algorithm with the initial learning rate of $0.03$, which gradually decreases with cosine annealing \cite{cosine}. We use different hyper-parameter settings of weight decay $(5e^{-2}$ to $7e^{-5})$, and momentum $(0.9$ to $0.999)$ for different datasets. Due to the scarcity of training data samples, we train the network for a large number of epochs, for example, for the retina segmentation task, the network is trained for $3000$ epochs and $200$ epochs for other tasks with batch size of $8$ and $16$.  We also use the Probability Correction Strategy (PCS) \cite{sarnet} during inference to improve the final prediction. The detail of Probability Correction Strategy (PCS) have been provided in Appendix \ref{sec:PCS}.\vspace{-1.5em}

\subsection{Evaluation Metric}\label{subsec:metric}\vspace{-0.5em}
We evaluate the performance of our best model using standard medical image segmentation metrics, i.e. mean dice coefficient (\textit{mDice}), mean Intersection-over-Union (\textit{mIoU}), and mean F1 (\textit{mF1}) scores. \vspace{-1.0em}
\subsection{Quantitative Results}\label{sec:quantitative-results}\vspace{-0.5em} We have evaluated TransResNet on three different segmentation tasks as discussed in Sec. \ref{subsec:datasets}, to demonstrate its effectiveness. We have compared our method with six SOTA methods for skin lesion segmentation, four for retinal vessel segmentation, and seven for polyp segmentation tasks.

\noindent \textbf{Results of Skin lesion Segmentation:}
We report our results for the skin lesion segmentation task on two datasets and compare them with six SOTA methods. Table \ref{table:skin-results}, shows that our model has achieved the highest performance on both validation ISIC-2016, and test-PH2 datasets on both evaluation metrics. Since these datasets have higher resolution images, the model has captured rich semantic information, resulting in enhanced performance. In addition, the samples from the PH2 dataset are not part of the training phase, which indicates that our model has better generalization ability and robustness towards skin lesion segmentation task than other SOTA approaches.

\noindent \textbf{Results of Retinal Vessel Segmentation:} We compare TransResNet with four SOTA methods on three high-resolution fundus imaging datasets for the retinal vessel segmentation task. Table \ref{table:retinal-results}, demonstrates that our method has surpassed all the other SOTA methods by a margin of $1.4\%$, $1.0\%$, and $0.9\%$ in terms of mean F1 (\textit{mF1}) score on HRF, IOSTAR, and CHASE\_DB1 datasets respectively, without applying any pre-processing technique on these datasets.\vspace{-0.5em} 

\begin{table}[!ht]
    \begin{minipage}{0.49\textwidth}
    \centering\small
\setlength{\tabcolsep}{4pt}
\setlength{\extrarowheight}{0.4pt}
        \scalebox{0.8}[0.8]{
\begin{tabular}{l|c|c|c|c}
\hline
\hline
\rowcolor{gray}
\textbf{Methods} &\multicolumn{2}{c|}{\textbf{ISIC-2016}}&\multicolumn{2}{c}{\textbf{test-{PH2}}}\\
\hline
\hline
& \textit{mIoU}$\uparrow$ & \textit{mDice}$\uparrow$ & \textit{mIoU}$\uparrow$ & \textit{mDice}$\uparrow$\\
\hline
\hline
\textbf{U-Net} \cite{unet} &0.825 &0.878 &0.739 &0.836\\
\textbf{U-Net++} \cite{unet++} &0.818 &0.889 &0.812 &0.889\\
\textbf{Attn U-Net} \cite{attentionunet} &0.797 &0.874 &0.695 &0.805\\
\textbf{CE-Net} \cite{cenet} & \cellcolor{green!15}\textbf{{0.842}} & \cellcolor{green!15}\textbf{{0.905}} & \cellcolor{green!15}\textbf{{0.824}} &0.894\\
\textbf{CA-Net} \cite{canet} &0.807 &0.881&0.751 &0.846 \\
\textbf{TransFuse} \cite{zhang2021transfuse} &0.840 &0.900 &0.823 &\cellcolor{green!15}\textbf{{0.897}}\\
\hline
\hline
\textbf{TransResNet} &\cellcolor{red!20}\textbf{0.843}&\cellcolor{red!20}\textbf{{0.907}} &\cellcolor{red!20}\textbf{{0.831}} &\cellcolor{red!20}\textbf{{0.905}}\\
\hline
\end{tabular}}
\caption{\footnotesize{Quantitative results on skin lesion segmentation datasets compared with six SOTA methods. The red and green color cells represent the highest and the second highest scores respectively. Performance is measured by mean Dice and mean IoU scores.}}\label{table:skin-results}
\end{minipage}
    \hfill
    \begin{minipage}{0.49\textwidth}
        \centering\small
\setlength{\tabcolsep}{8pt}
\setlength{\extrarowheight}{4.0pt}
        \scalebox{0.8}[0.8]{
\begin{tabular}{l|c|c|c}
%column{3} = {teal7},
%cell{2}{3} = {yellow7},
\hline
\hline
\rowcolor{gray}
\textbf{Methods} &\textbf{HRF} &\textbf{IOSTAR}& \textbf{CHASE}\\
\hline
\hline
& \textit{mF1}$\uparrow$ & \textit{mF1}$\uparrow$ & \textit{mF1}$\uparrow$\\
\hline
\hline
\textbf{DRIU} \cite{driu}&0.783 &\cellcolor{green!15}\textbf{0.825} &0.810\\
\textbf{HED} \cite{hed}&0.783 &\cellcolor{green!15}\textbf{0.825} &0.810\\
\textbf{M2U-Net} \cite{m2u} &0.780 &0.817 &0.802\\
\textbf{U-Net} \cite{unet} &\cellcolor{green!15}\textbf{0.788} &0.812 &\cellcolor{green!15}\textbf{0.812} \\
\hline
\hline
\textbf{TransResNet} &\cellcolor{red!20}\textbf{0.802} &\cellcolor{red!20}\textbf{0.835} &\cellcolor{red!20}\textbf{0.821}\\
\hline
\end{tabular}}
\caption{\footnotesize{Quantitative results on retinal vessel segmentation datasets compared with four SOTA methods. The red and green color cells represent the highest and the second highest scores respectively. Performance is measured by the mean F1 score.}}\label{table:retinal-results}
\end{minipage}
\end{table}\vspace{-1.0em}
\noindent \textbf{Results of Polyp Segmentation:}
The performance of TransResNet for polyp segmentation has been evaluated and compared with seven SOTA methods across five different benchmark datasets. The quantitative results are shown in Table \ref{table:polyp-results}. As highlighted from the scores, our proposed architecture did not surpass some SOTA methods on polyp segmentation tasks except for the ClinicDB dataset. In addition, the mean dice scores of our method on Kvasir and EndoScence datasets are also very close to the highest and second highest SOTA methods, i.e., SANet, and TransFuse, with a minor difference of $0.037$ on Kvasir, and $0.028$ on ColonDB datasets respectively. We also find that our method performs unfavorably and does not generalize on ColonDB and ETIS datasets because these datasets have lower image resolution. Our method is better suited for high resolution images. We have provided detailed information and analysis in Appendix \ref{sec:datasets} about datasets.

\begin{table}[!ht]
\begin{minipage}{0.59\textwidth}
\noindent  \textbf{Model Performance on different Image Resolutions:} From the segmentation results of our proposed method, we have analyzed four cases regarding the model performance based on input images during training and inference, which is summarized in Table \ref{table:img_res_analyses}. Given the results, one can conclude that image resolution is a major issue in the context of  model performance. Our model design is highly suitable for high resolution images.
\end{minipage}
    \hfill
    \begin{minipage}{0.39\textwidth}
\centering\small
\setlength{\tabcolsep}{3pt}
\scalebox{0.65}[0.65]{
\begin{tabular}{l |l|l} %c} % centered columns (4 columns)
\hline
\hline%inserts double horizontal lines
% Momentum & Mean $\pm$ std & Max\ \\ [0.1ex] % inserts table
\rowcolor{gray!50}
\pbox{\textwidth}{\textbf{Train Image} \\ \textbf{Resolution} \vphantom{g}} & \pbox{\textwidth}{\textbf{Test Image} \\ \textbf{Resolution} \vphantom{g}} & \pbox{\textwidth}{\textbf{Model } \\ \textbf{Performance} \vphantom{g}}\\    
%heading
\hline % inserts single horizontal line
\hline
\pbox{\textwidth}{Higher \vphantom{g}} & \pbox{\textwidth}{Higher \vphantom{g}} & \pbox{\textwidth}{Increases \vphantom{g}}\\

\pbox{\textwidth}{Lower \& upscaled \vphantom{g}} & \pbox{\textwidth}{Higher \vphantom{g}} & \pbox{\textwidth}{Decreases \vphantom{g}}\\

\pbox{\textwidth}{Higher \vphantom{g}} & \pbox{\textwidth}{Lower \& upscaled \vphantom{g}} & \pbox{\textwidth}{Increases \vphantom{g}}\\

\pbox{\textwidth}{Lower \& upscaled \vphantom{g}} & \pbox{\textwidth}{Lower \& upscaled \vphantom{g}} & \pbox{\textwidth}{Slightly decreases \vphantom{g}}\\
\hline
\end{tabular}}
\caption{\footnotesize{Analysis of model performance based on the image resolution during training and inference. Lower resolution images are upscaled to $1024\times1024$.}}\label{table:img_res_analyses}
\end{minipage}
\end{table}

\begin{table*}[!h]
\centering
\vspace{-0.5em}
\caption{\footnotesize{Quantitative results on polyp segmentation datasets compared with seven SOTA methods. The red and green color cells represent the highest and the second highest scores respectively. Performance is measured by mean Dice and mean IoU scores. "-" indicates results are not available.}}\label{table:polyp-results}\vspace{0.5em}
\scalebox{0.72}{
\begin{tabular}{l|c|c|c|c|c|c|c|c|c|c}
%column{3} = {teal7},
%cell{2}{3} = {yellow7},
\hline
\hline
\rowcolor{gray}
\textbf{Methods} &\multicolumn{2}{c|}{\textbf{Kvasir}}&\multicolumn{2}{c|}{\textbf{ClinicDB}}&\multicolumn{2}{c|}{\textbf{ColonDB}}&\multicolumn{2}{c|}{\textbf{EndoScene}}& \multicolumn{2}{c}{\textbf{ETIS}}\\
\hline
\hline
& \textit{mDice}$\uparrow$ & \textit{mIoU}$\uparrow$ & \textit{mDice}$\uparrow$ & \textit{mIoU}$\uparrow$ & \textit{mDice}$\uparrow$& \textit{mIoU}$\uparrow$ & \textit{mDice}$\uparrow$ & \textit{mIoU}$\uparrow$ & \textit{mDice}$\uparrow$ & \textit{mIoU}$\uparrow$\\
\hline
\hline
\textbf{U-Net} \cite{unet} &0.818 &0.746 &0.823 &0.750& 0.512& 0.444& 0.710& 0.627& 0.398& 0.335\\
\textbf{U-Net++} \cite{unet++} &0.821 &0.743 &0.794 &0.729 &0.483 &0.410 &0.707 &0.624& 0.401 &0.344\\
\textbf{ResUNet++} \cite{recurrentunet} &0.813 &0.793 &0.796 &0.796 &- &- &- &- &- &- \\
\textbf{SFA} \cite{sfa}&0.723 &0.611 &0.700 &0.607 &0.469 &0.347 &0.467 &0.329& 0.297 &0.217\\ 
\textbf{PraNet} \cite{pranet} &0.898 &0.840 &0.899 &0.849 &0.712 &0.640 &0.871 &0.797& 0.628 &0.567 \\
\textbf{SANet} \cite{sarnet} &\cellcolor{green!15}\textbf{0.904} &\cellcolor{green!15}\textbf{0.847} &0.916 &0.859 &\cellcolor{green!15}\textbf{0.753} &\cellcolor{green!15}\textbf{0.670} &\cellcolor{green!15}\textbf{0.888}
&\cellcolor{green!15}\textbf{0.815}
&\cellcolor{red!20}\textbf{0.750} &\cellcolor{green!15}\textbf{0.654}\\
\textbf{TransFuse} \cite{zhang2021transfuse} &\cellcolor{red!20}\textbf{0.918} &\cellcolor{red!20}\textbf{0.868} &\cellcolor{red!20}\textbf{0.918} &\cellcolor{red!20}\textbf{0.868} &\cellcolor{red!20}\textbf{0.773} &\cellcolor{red!20}\textbf{0.696}
&\cellcolor{red!20}\textbf{0.902} &\cellcolor{red!20}\textbf{0.833} &\cellcolor{green!15}\textbf{0.733} &\cellcolor{red!20}\textbf{0.659}\\
\hline
\hline
\textbf{TransResNet}&0.881 &0.824 &\cellcolor{green!15}\textbf{0.917} &\cellcolor{green!15}\textbf{0.861} &0.685 &0.604 &0.874 &0.804 &0.564 &0.493\\
\hline
\end{tabular}
}
\end{table*}\vspace{-1.5em}

\subsection{Qualitative Results}\label{sec:qualitative-results}
\begin{figure*}[!ht]
\begin{minipage}{0.5\textwidth}
\centering
\begin{minipage}{0.24\textwidth}
      \centering
       % lelf lower right up trim= 7.5mm 0mm 0mm 10mm
    \includegraphics[height=1.2cm, width=\linewidth]{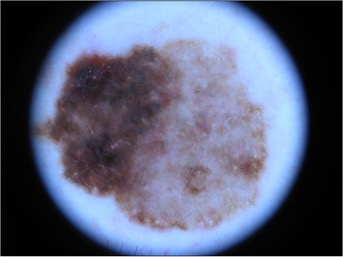}
  \end{minipage}
  \begin{minipage}{0.24\textwidth}
      \centering
       % lelf lower right up trim= 7.5mm 0mm 0mm 10mm
   \includegraphics[height=1.2cm, width=\linewidth]{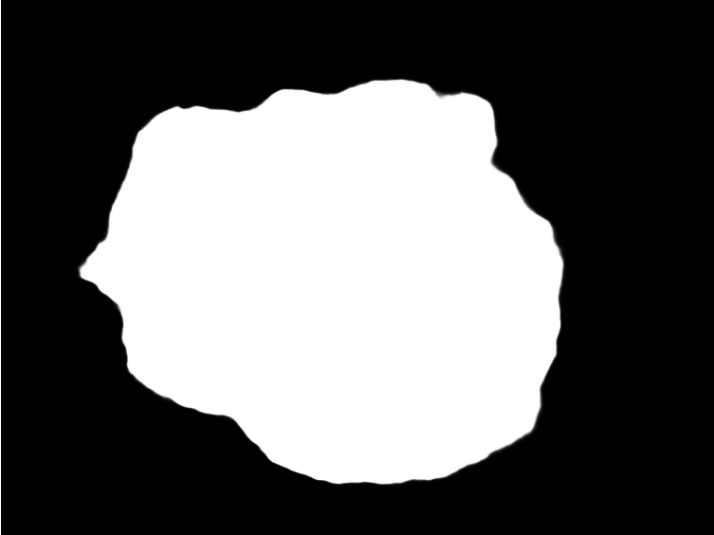}
  \end{minipage}
\begin{minipage}{0.24\textwidth}
      \centering
       % lelf lower right up
  \includegraphics[height=1.2cm, width=\linewidth]{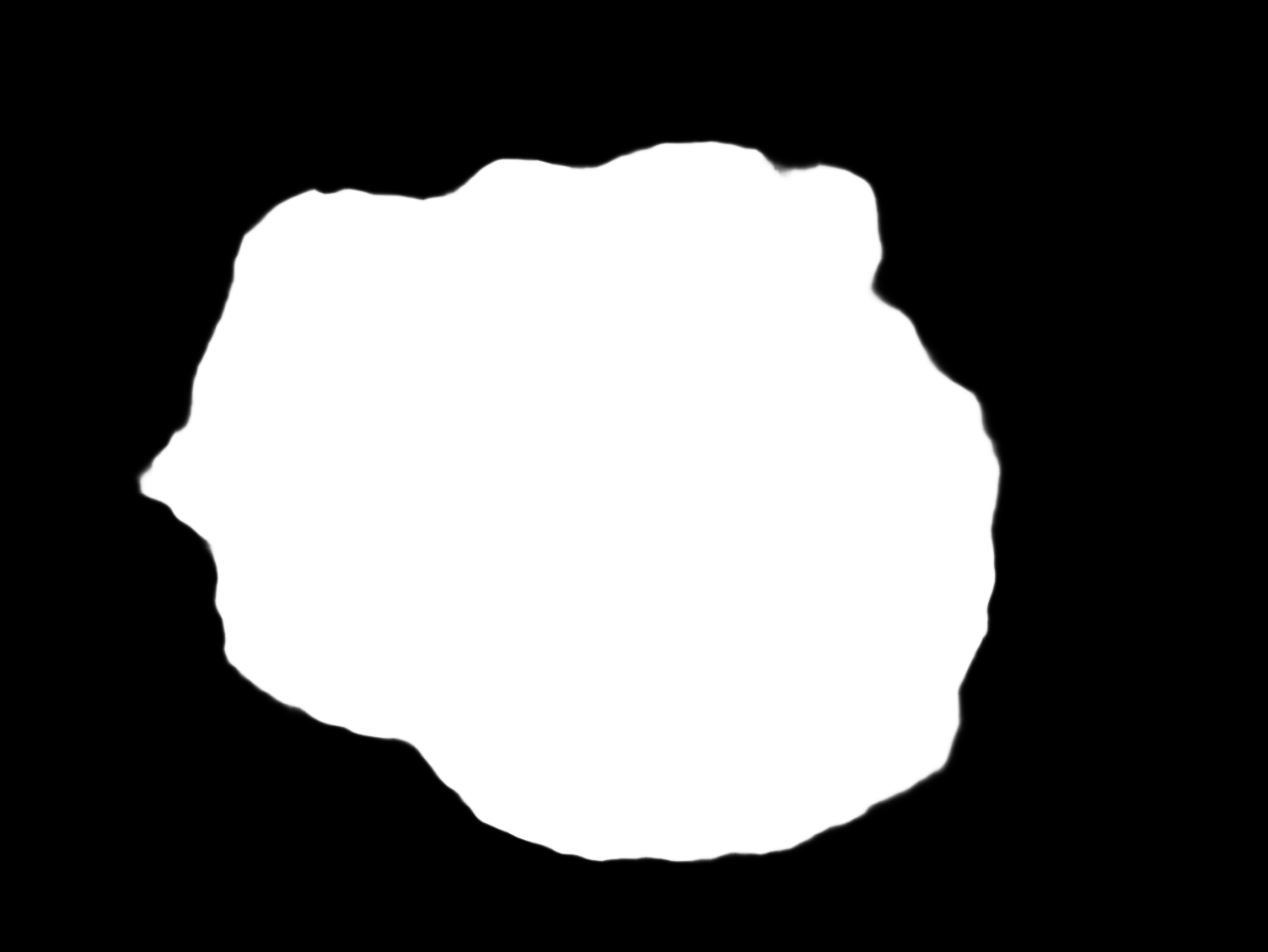}
  \end{minipage}
  \\
  \begin{minipage}{0.24\textwidth}
      \centering
       % lelf lower right up trim= 7.5mm 0mm 0mm 10mm
    \includegraphics[height=1.2cm, width=\linewidth]{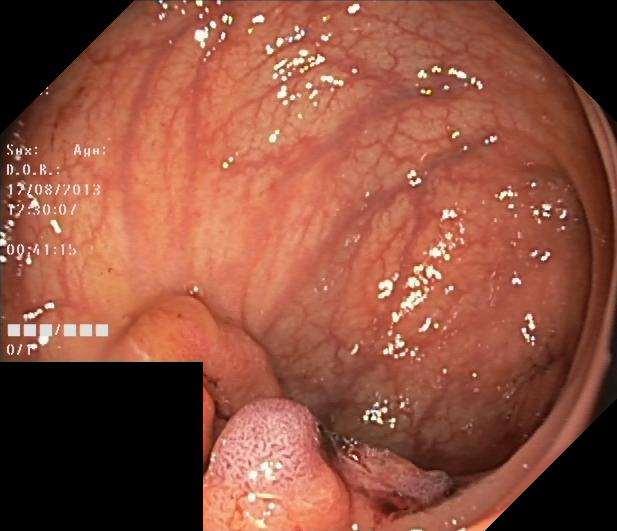}
  \end{minipage}
  \begin{minipage}{0.24\textwidth}
      \centering
       % lelf lower right up trim= 7.5mm 0mm 0mm 10mm
   \includegraphics[height=1.2cm, width=\linewidth]{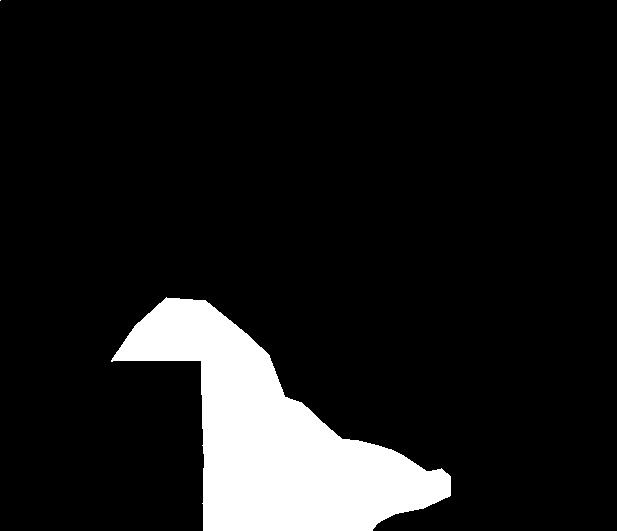}
  \end{minipage}
\begin{minipage}{0.24\textwidth}
      \centering
       % lelf lower right up
  \includegraphics[height=1.2cm, width=\linewidth]{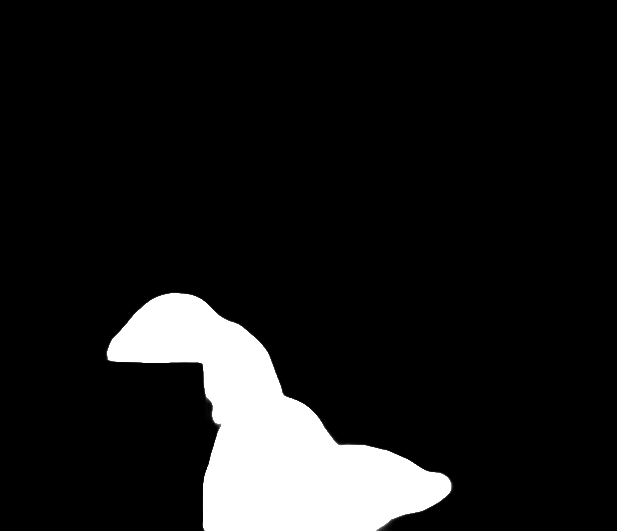}
  \end{minipage}
  \\
  \begin{minipage}{0.24\textwidth}
      \centering
       % lelf lower right up trim= 7.5mm 0mm 0mm 10mm
    \includegraphics[height=1.2cm, width=\linewidth]{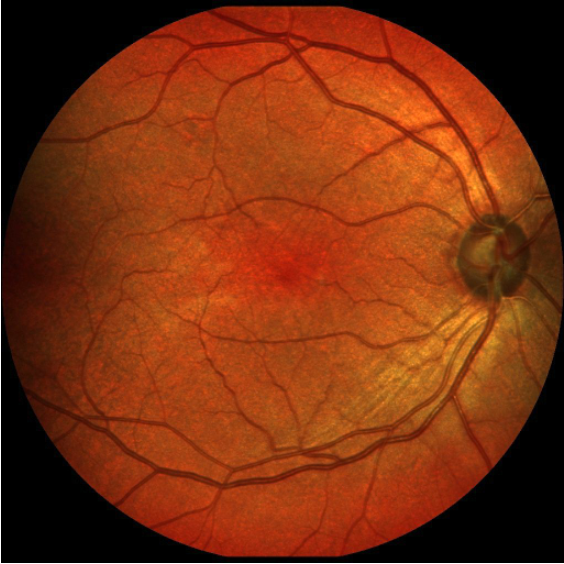}
    \tiny \textbf{IMAGES}
  \end{minipage}
  \begin{minipage}{0.24\textwidth}
      \centering
       % lelf lower right up trim= 7.5mm 0mm 0mm 10mm
   \includegraphics[height=1.2cm, width=\linewidth]{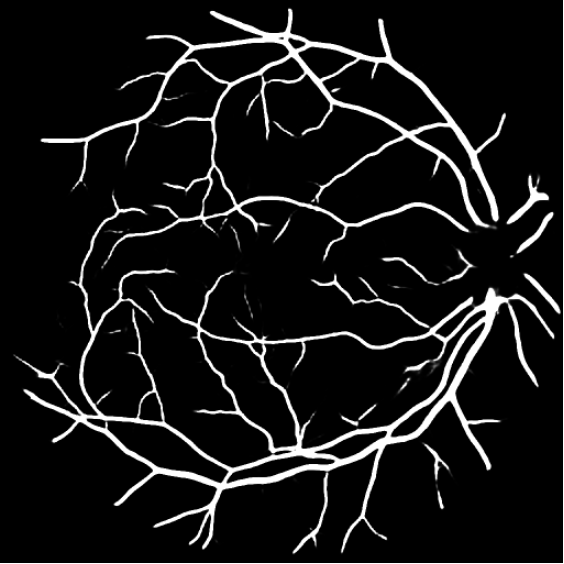}
   \tiny \textbf{GT}
  \end{minipage}
\begin{minipage}{0.24\textwidth}
      \centering
       % lelf lower right up
  \includegraphics[height=1.2cm, width=\linewidth]{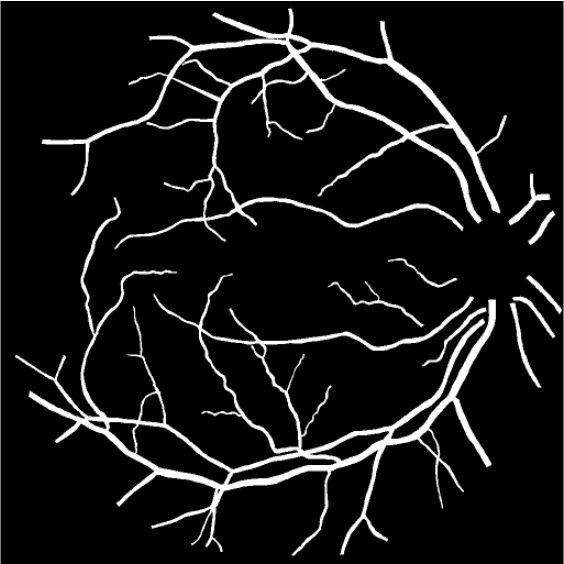}
  \tiny \textbf{PRED.}
  \end{minipage}

  \end{minipage}
  \hfill
\begin{minipage}{0.5\textwidth}

\centering
\begin{minipage}{0.24\textwidth}
      \centering
       % lelf lower right up trim= 7.5mm 0mm 0mm 10mm
    \includegraphics[height=1.2cm, width=\linewidth]{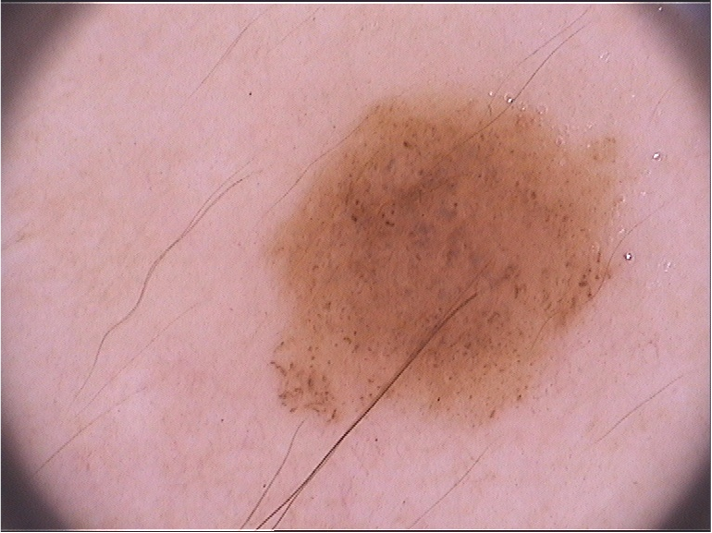}
  \end{minipage}
  \begin{minipage}{0.24\textwidth}
      \centering
       % lelf lower right up trim= 7.5mm 0mm 0mm 10mm
   \includegraphics[height=1.2cm, width=\linewidth]{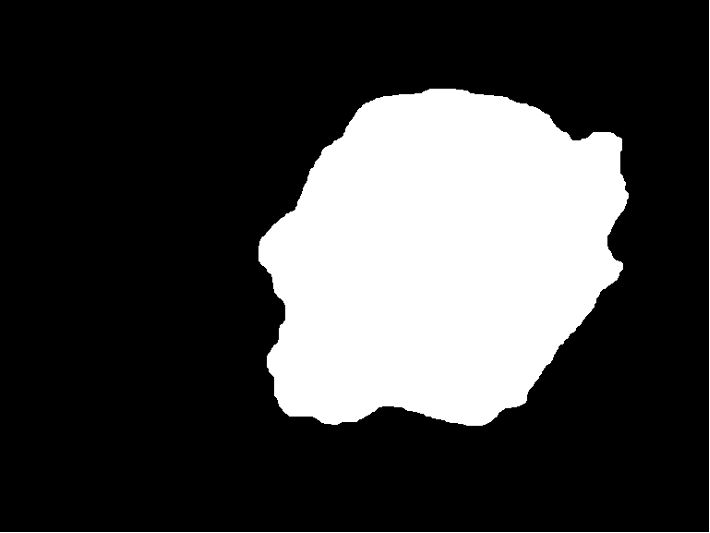}
  \end{minipage}
\begin{minipage}{0.24\textwidth}
      \centering
       % lelf lower right up
  \includegraphics[height=1.2cm, width=\linewidth]{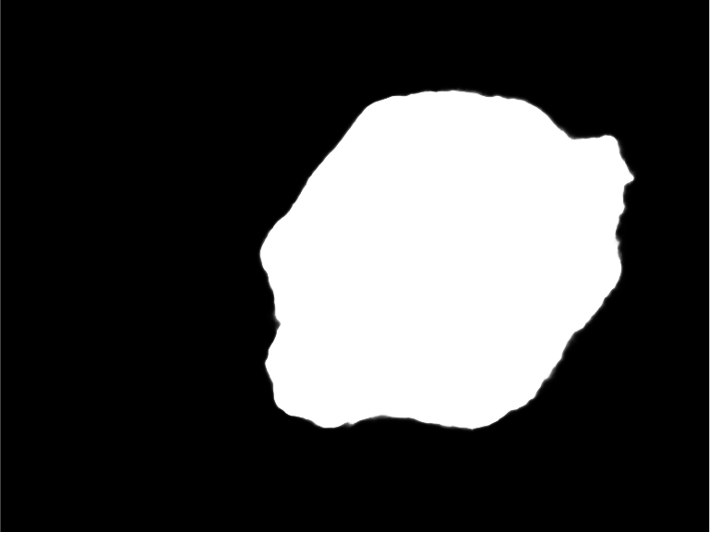}
  \end{minipage}
  \\
  \begin{minipage}{0.24\textwidth}
      \centering
       % lelf lower right up trim= 7.5mm 0mm 0mm 10mm
    \includegraphics[height=1.2cm, width=\linewidth]{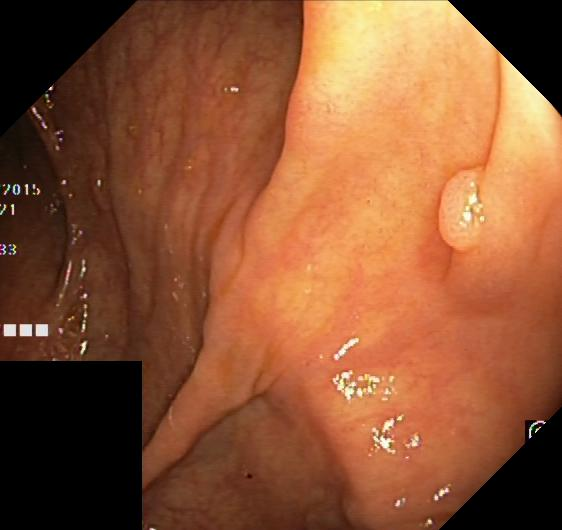}
  \end{minipage}
  \begin{minipage}{0.24\textwidth}
      \centering
       % lelf lower right up trim= 7.5mm 0mm 0mm 10mm
   \includegraphics[height=1.2cm, width=\linewidth]{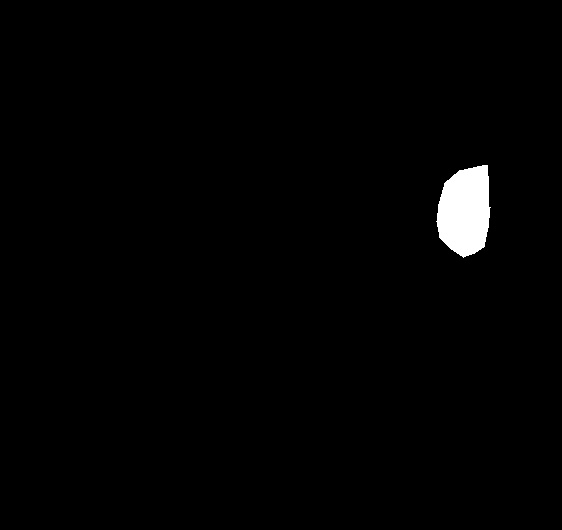}
  \end{minipage}
\begin{minipage}{0.24\textwidth}
      \centering
       % lelf lower right up
  \includegraphics[height=1.2cm, width=\linewidth]{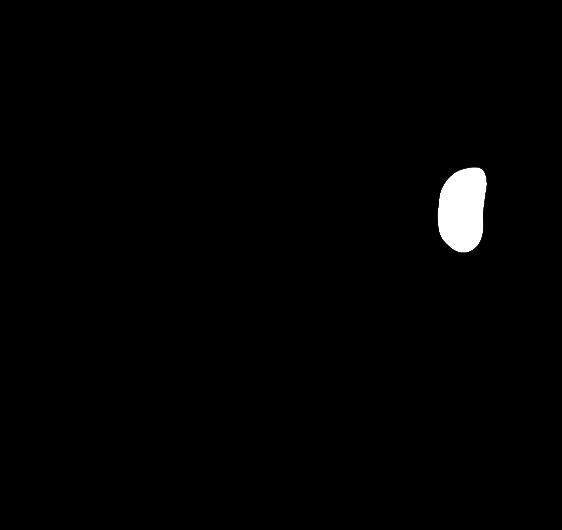}
  \end{minipage}
  \\
  \begin{minipage}{0.24\textwidth}
      \centering
       % lelf lower right up trim= 7.5mm 0mm 0mm 10mm
    \includegraphics[height=1.2cm, width=\linewidth]{images/Visualizations/retinal_case4_img.png}
    \tiny \textbf{IMAGES}
  \end{minipage}
  \begin{minipage}{0.24\textwidth}
      \centering
       % lelf lower right up trim= 7.5mm 0mm 0mm 10mm
   \includegraphics[height=1.2cm, width=\linewidth]{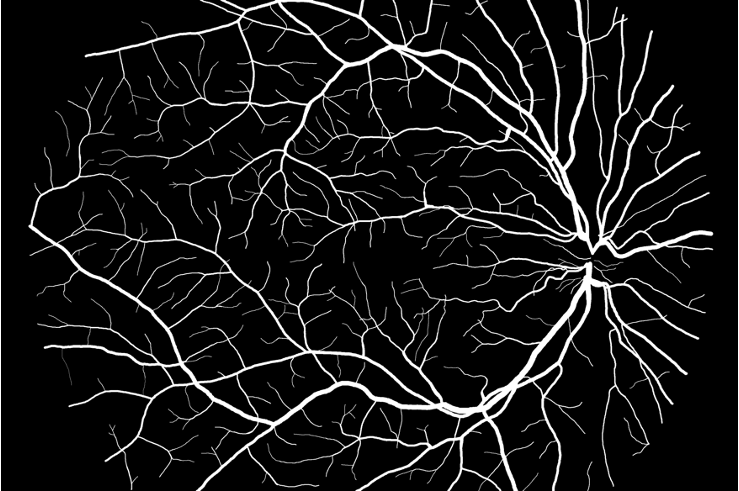}
   \tiny \textbf{GT}
  \end{minipage}
\begin{minipage}{0.24\textwidth}
      \centering
       % lelf lower right up
  \includegraphics[height=1.2cm, width=\linewidth]{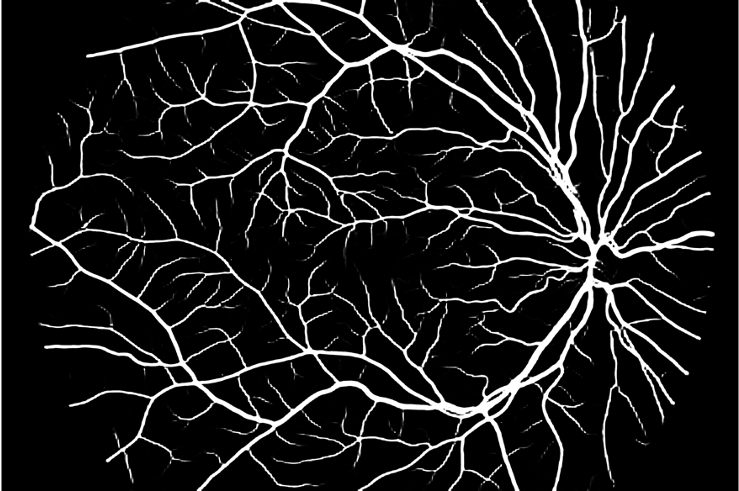}
  \tiny \textbf{PRED.}
  \end{minipage}
  \end{minipage}
  \vfill
  \begin{minipage}{0.96\textwidth}
\caption{ \footnotesize{Qualitative results on all three segmentation tasks. The figure shows an example image, ground truth (GT) and predicted (PRED) segmentation mask for the skin lesion segmentation task (row 1), the polyp segmentation task (row 2) and retinal vessel segmentation task (row 3). }}\label{fig:visualizations}
  \end{minipage}
\vspace{-1em}
\end{figure*}

\vspace{-0.5em}
The Fig. \ref{fig:visualizations} shows predicted segmentation masks for some input images for each segmentation task. From the predicted masks, we can easily conclude that our method has not only performed an accurate prediction but also suppressed background noise. Additional visual results have been provided in Appendix \ref{sec:add-visual-results}.\vspace{-0.5em}

\vspace{-0.75em}

\subsection{Ablation Study}\label{sec:ablation}\vspace{-0.5em}
In order to evaluate the effectiveness of the proposed method, we analyze our approach using two ablative studies, which are as follows: \textbf{(a)} Effect of network performance by using different feature map pairs in Cross Grafting Module (CGM) and \textbf{(b)} Effect of network performance by eliminating each module from the proposed architecture. 

\noindent\textbf{Ablation Study for Grafted Features:} For better  design of Cross Grafting Module (CGM), we conduct an ablation study, in which we change the feature map pairs used in the grafted module and study its impact on the performance of the network. In Table \ref{table:ablation}, we present the quantitative results for two datasets: ISIC-2016 and PH2. From the Table \ref{table:ablation}, we observe that the performance gradually increases and then tends to decrease, indicating that the pair ($\textbf{R}_{f_{5}}$, $\textbf{S}_{f_{2}}$) is suitable for grafting. The main reason is that the spatial sizes of both feature maps are very close, and the information captured by each model corresponds to each other, increasing the network performance through grafting.

\noindent\textbf{Ablation Study for Elimination each Module:} The proposed framework comprises of Swin-B transformer, Resnet-18 backbone, and Cross Grafting Module (CGM) that are used for learning both local and global features. To experimentally evaluate the effect and contribution of each module in the generalization performance, we selectively remove one of the modules, as shown in Table \ref{table:ablation-study}. The qualitative finding suggests that removing any of the modules from the architecture results in a performance degradation.\vspace{-0.5em}

\begin{table}[!ht]
     \begin{minipage}{.49\textwidth}
        \centering\small
        \setlength{\tabcolsep}{4pt}

        \scalebox{0.65}[0.65]{
\begin{tabular}{l|c|c|c|c|c}
%column{3} = {teal7},
%cell{2}{3} = {yellow7},
\hline
\hline
\rowcolor{gray}
\textbf{Feature Pairs} & \textbf{Spatial Dimensions} &\multicolumn{2}{c|}{\textbf{val-ISIC-2016}}&\multicolumn{2}{c}{\textbf{test-PH2}}\\
\hline
\hline
& & \textit{mIoU}$\uparrow$ & \textit{mDice}$\uparrow$ & \textit{mIoU}$\uparrow$ & \textit{mDice}$\uparrow$\\
\hline
\hline
$(\textbf{R}_{f_{5}}$ , $\textbf{S}_{f_{1}})$&(32$\times$32 , 56$\times$56)&0.829&0.881&0.824&0.881\\
$(\textbf{R}_{f_{5}}$ , $\textbf{S}_{f_{2}})$&(32$\times$32 , 28$\times$28)&\textbf{0.843}&\textbf{0.907}&\textbf{0.831}&\textbf{0.905}\\
$(\textbf{R}_{f_{5}}$ , $\textbf{S}_{f_{3}})$&(32$\times$32 ,
14$\times$14)&0.836&0.892&0.825&0.896\\
$(\textbf{R}_{f_{5}}$ , $\textbf{S}_{f_{4}})$&(32$\times$32 , 14$\times$ 14)&0.819&0.862&0.812&0.874  \\
\hline
\end{tabular}}
\caption{\footnotesize{An ablation study for TransResNet on ISIC-2016, and PH2 datasets for selection of feature pairs from Swin-transformer and ResNet for grafting. }\label{table:ablation}}
\end{minipage}
    \hfill
     \begin{minipage}{.48\textwidth}
        \centering\small
        \setlength{\tabcolsep}{4pt}

        \scalebox{0.65}[0.65]{
\begin{tabular}{l|c|c|c|c|c|c}
\hline
\hline
\rowcolor{gray}
\textbf{ResNet-18}&\textbf{Swin-B}&\textbf{CGM}&\multicolumn{2}{c|}{\textbf{val-ISIC-2016}}&\multicolumn{2}{c}{\textbf{test-{PH2}}}\\
\hline
\hline
\multicolumn{3}{c|}{} &\textit{mDice}$\uparrow$ & \textit{mIoU}$\uparrow$ & \textit{mDice}$\uparrow$ & \textit{mIoU}$\uparrow$\\
\hline
\hline
\cmark & \xmark & \xmark & 0.879 & 0.819 & 0.879 & 0.814\\
\xmark & \cmark & \xmark & 0.881 & 0.821 & 0.884 & 0.806\\
\cmark & \cmark & \xmark & 0.889 & 0.832 & 0.900 & 0.821\\
\cmark & \cmark & \cmark & \textbf{0.907} & \textbf{0.843} & \textbf{0.905} & \textbf{0.831}\\
\hline
\end{tabular}}
\caption{\footnotesize{An ablation study to analyze the effect on the overall performance of the proposed method by eliminating each module. }\label{table:ablation-study}}
\end{minipage}\vspace{-1.0em}
\end{table}\vspace{-1em}

\section{Conclusion}\label{sec:conclusion}\vspace{-0.5em}
In this paper, we present the TransResNet architecture for the segmentation of high-resolution medical images. A key component of TransResNet is the \emph{Cross Grafting Module (CGM)}, which is used to learn \textit{grafted features} with rich semantic and global information, allowing accurate prediction of segmentation masks during decoding. Extensive experiments on ten different datasets for three medical segmentation tasks indicate that our architecture shows better results on high-resolution images. One of the main limitations of our architecture, is that it is computationally expensive. With our research work, we intend to introduce the scientific community with AI-based model for high-resolution medical image segmentation. This will open new directions for conducting research on this problem as the demand for learning-based models with capability of efficiently processing the high-resolution images is expected to rise. Future direction in this line of research includes extending the proposed method to multi-class medical image segmentation and making it computationally less expensive.

\bibliography{egbib}

\begin{thebibliography}{47}
\providecommand{\natexlab}[1]{#1}
\providecommand{\url}[1]{\texttt{#1}}
\expandafter\ifx\csname urlstyle\endcsname\relax
  \providecommand{\doi}[1]{doi: #1}\else
  \providecommand{\doi}{doi: \begingroup \urlstyle{rm}\Url}\fi

\bibitem[Alom et~al.(2018)Alom, Hasan, Yakopcic, Taha, and Asari]{recurrentunet}
Md~Zahangir Alom, Mahmudul Hasan, Chris Yakopcic, Tarek~M Taha, and Vijayan~K Asari.
\newblock Recurrent residual convolutional neural network based on u-net (r2u-net) for medical image segmentation.
\newblock \emph{arXiv preprint arXiv:1802.06955}, 2018.

\bibitem[Asgari~Taghanaki et~al.(2021)Asgari~Taghanaki, Abhishek, Cohen, Cohen-Adad, and Hamarneh]{asgari2021deep}
Saeid Asgari~Taghanaki, Kumar Abhishek, Joseph~Paul Cohen, Julien Cohen-Adad, and Ghassan Hamarneh.
\newblock Deep semantic segmentation of natural and medical images: a review.
\newblock \emph{Artificial Intelligence Review}, 54\penalty0 (1):\penalty0 137--178, 2021.

\bibitem[Bernal et~al.(2015)Bernal, S{\'a}nchez, Fern{\'a}ndez-Esparrach, Gil, Rodr{\'\i}guez, and Vilari{\~n}o]{clinicdb}
Jorge Bernal, F~Javier S{\'a}nchez, Gloria Fern{\'a}ndez-Esparrach, Debora Gil, Cristina Rodr{\'\i}guez, and Fernando Vilari{\~n}o.
\newblock Wm-dova maps for accurate polyp highlighting in colonoscopy: Validation vs. saliency maps from physicians.
\newblock \emph{Computerized medical imaging and graphics}, 43:\penalty0 99--111, 2015.

\bibitem[Chen et~al.(2021)Chen, Lu, Yu, Luo, Adeli, Wang, Lu, Yuille, and Zhou]{chen2021transunet}
Jieneng Chen, Yongyi Lu, Qihang Yu, Xiangde Luo, Ehsan Adeli, Yan Wang, Le~Lu, Alan~L Yuille, and Yuyin Zhou.
\newblock Transunet: Transformers make strong encoders for medical image segmentation.
\newblock \emph{arXiv preprint arXiv:2102.04306}, 2021.

\bibitem[Dong et~al.(2019)Dong, Xu, Liang, Jiang, Dai, and Xing]{dong2019neural}
Nanqing Dong, Min Xu, Xiaodan Liang, Yiliang Jiang, Wei Dai, and Eric Xing.
\newblock Neural architecture search for adversarial medical image segmentation.
\newblock In \emph{International Conference on Medical Image Computing and Computer-Assisted Intervention}, pages 828--836. Springer, 2019.

\bibitem[Dosovitskiy et~al.(2020)Dosovitskiy, Beyer, Kolesnikov, Weissenborn, Zhai, Unterthiner, Dehghani, Minderer, Heigold, Gelly, et~al.]{dosovitskiy2020image}
Alexey Dosovitskiy, Lucas Beyer, Alexander Kolesnikov, Dirk Weissenborn, Xiaohua Zhai, Thomas Unterthiner, Mostafa Dehghani, Matthias Minderer, Georg Heigold, Sylvain Gelly, et~al.
\newblock An image is worth 16x16 words: Transformers for image recognition at scale.
\newblock \emph{arXiv preprint arXiv:2010.11929}, 2020.

\bibitem[Fan et~al.(2020)Fan, Ji, Zhou, Chen, Fu, Shen, and Shao]{pranet}
Deng-Ping Fan, Ge-Peng Ji, Tao Zhou, Geng Chen, Huazhu Fu, Jianbing Shen, and Ling Shao.
\newblock Pranet: Parallel reverse attention network for polyp segmentation.
\newblock In \emph{International conference on medical image computing and computer-assisted intervention}, pages 263--273. Springer, 2020.

\bibitem[Fang et~al.(2019)Fang, Chen, Yuan, and Tong]{sfa}
Yuqi Fang, Cheng Chen, Yixuan Yuan, and Kai-yu Tong.
\newblock Selective feature aggregation network with area-boundary constraints for polyp segmentation.
\newblock In \emph{International Conference on Medical Image Computing and Computer-Assisted Intervention}, pages 302--310. Springer, 2019.

\bibitem[{Fraz} et~al.(2012){Fraz}, {Remagnino}, {Hoppe}, {Uyyanonvara}, {Rudnicka}, {Owen}, and {Barman}]{cashdb1}
M.~M. {Fraz}, P.~{Remagnino}, A.~{Hoppe}, B.~{Uyyanonvara}, A.~R. {Rudnicka}, C.~G. {Owen}, and S.~A. {Barman}.
\newblock An ensemble classification-based approach applied to retinal blood vessel segmentation.
\newblock \emph{IEEE Transactions on Biomedical Engineering}, 59\penalty0 (9):\penalty0 2538--2548, Sep. 2012.
\newblock ISSN 0018-9294.
\newblock \doi{10.1109/TBME.2012.2205687}.

\bibitem[Fraz et~al.(2012)Fraz, Remagnino, Hoppe, Uyyanonvara, Rudnicka, Owen, and Barman]{fensemble}
Muhammad~Moazam Fraz, Paolo Remagnino, Andreas Hoppe, Bunyarit Uyyanonvara, Alicja~R Rudnicka, Christopher~G Owen, and Sarah~A Barman.
\newblock An ensemble classification-based approach applied to retinal blood vessel segmentation.
\newblock \emph{IEEE Transactions on Biomedical Engineering}, 59\penalty0 (9):\penalty0 2538--2548, 2012.

\bibitem[Gibson et~al.(2018)Gibson, Giganti, Hu, Bonmati, Bandula, Gurusamy, Davidson, Pereira, Clarkson, and Barratt]{gibson2018organ}
Eli Gibson, Francesco Giganti, Yipeng Hu, Ester Bonmati, Steve Bandula, Kurinchi Gurusamy, Brian Davidson, Stephen~P Pereira, Matthew~J Clarkson, and Dean~C Barratt.
\newblock Automatic multi-organ segmentation on abdominal ct with dense v-networks.
\newblock \emph{IEEE transactions on medical imaging}, 37\penalty0 (8):\penalty0 1822--1834, 2018.

\bibitem[Gu et~al.(2020)Gu, Wang, Song, Huang, Aertsen, Deprest, Ourselin, Vercauteren, and Zhang]{canet}
Ran Gu, Guotai Wang, Tao Song, Rui Huang, Michael Aertsen, Jan Deprest, S{\'e}bastien Ourselin, Tom Vercauteren, and Shaoting Zhang.
\newblock Ca-net: Comprehensive attention convolutional neural networks for explainable medical image segmentation.
\newblock \emph{IEEE transactions on medical imaging}, 40\penalty0 (2):\penalty0 699--711, 2020.

\bibitem[Gu et~al.(2019)Gu, Cheng, Fu, Zhou, Hao, Zhao, Zhang, Gao, and Liu]{cenet}
Zaiwang Gu, Jun Cheng, Huazhu Fu, Kang Zhou, Huaying Hao, Yitian Zhao, Tianyang Zhang, Shenghua Gao, and Jiang Liu.
\newblock Ce-net: Context encoder network for 2d medical image segmentation.
\newblock \emph{IEEE transactions on medical imaging}, 38\penalty0 (10):\penalty0 2281--2292, 2019.

\bibitem[Gutman et~al.(2016)Gutman, Codella, Celebi, Helba, Marchetti, Mishra, and Halpern]{isic2016}
David Gutman, Noel~CF Codella, Emre Celebi, Brian Helba, Michael Marchetti, Nabin Mishra, and Allan Halpern.
\newblock Skin lesion analysis toward melanoma detection: A challenge at the international symposium on biomedical imaging (isbi) 2016, hosted by the international skin imaging collaboration (isic).
\newblock \emph{arXiv preprint arXiv:1605.01397}, 2016.

\bibitem[Hatamizadeh et~al.(2022)Hatamizadeh, Nath, Tang, Yang, Roth, and Xu]{hatamizadeh2022swin}
Ali Hatamizadeh, Vishwesh Nath, Yucheng Tang, Dong Yang, Holger Roth, and Daguang Xu.
\newblock Swin unetr: Swin transformers for semantic segmentation of brain tumors in mri images.
\newblock \emph{arXiv preprint arXiv:2201.01266}, 2022.

\bibitem[He et~al.(2016)He, Zhang, Ren, and Sun]{resnet}
Kaiming He, Xiangyu Zhang, Shaoqing Ren, and Jian Sun.
\newblock Deep residual learning for image recognition.
\newblock In \emph{Proceedings of the IEEE conference on computer vision and pattern recognition}, pages 770--778, 2016.

\bibitem[Isaac and Kulkarni(2015)]{super}
Jithin~Saji Isaac and Ramesh Kulkarni.
\newblock Super resolution techniques for medical image processing.
\newblock In \emph{2015 International Conference on Technologies for Sustainable Development (ICTSD)}. IEEE, 2015.

\bibitem[Jha et~al.(2020)Jha, Smedsrud, Riegler, Halvorsen, Lange, Johansen, and Johansen]{kvasir}
Debesh Jha, Pia~H Smedsrud, Michael~A Riegler, P{\aa}l Halvorsen, Thomas~de Lange, Dag Johansen, and H{\aa}vard~D Johansen.
\newblock Kvasir-seg: A segmented polyp dataset.
\newblock In \emph{International Conference on Multimedia Modeling}, pages 451--462. Springer, 2020.

\bibitem[Jin et~al.(2019)Jin, Meng, Pham, Chen, Wei, and Su]{dunet}
Qiangguo Jin, Zhaopeng Meng, Tuan~D Pham, Qi~Chen, Leyi Wei, and Ran Su.
\newblock Dunet: A deformable network for retinal vessel segmentation.
\newblock \emph{Knowledge-Based Systems}, 178:\penalty0 149--162, 2019.

\bibitem[Laibacher et~al.(2019)Laibacher, Weyde, and Jalali]{m2u}
Tim Laibacher, Tillman Weyde, and Sepehr Jalali.
\newblock M2u-net: Effective and efficient retinal vessel segmentation for real-world applications.
\newblock In \emph{Proceedings of the IEEE/CVF Conference on Computer Vision and Pattern Recognition Workshops}, pages 0--0, 2019.

\bibitem[Liu et~al.(2021)Liu, Lin, Cao, Hu, Wei, Zhang, Lin, and Guo]{swin}
Ze~Liu, Yutong Lin, Yue Cao, Han Hu, Yixuan Wei, Zheng Zhang, Stephen Lin, and Baining Guo.
\newblock Swin transformer: Hierarchical vision transformer using shifted windows.
\newblock In \emph{Proceedings of the IEEE/CVF International Conference on Computer Vision}, pages 10012--10022, 2021.

\bibitem[Loshchilov and Hutter(2016)]{cosine}
Ilya Loshchilov and Frank Hutter.
\newblock Sgdr: Stochastic gradient descent with warm restarts.
\newblock \emph{arXiv preprint arXiv:1608.03983}, 2016.

\bibitem[Maninis et~al.(2016)Maninis, Pont-Tuset, Arbel{\'a}ez, and Gool]{driu}
Kevis-Kokitsi Maninis, Jordi Pont-Tuset, Pablo Arbel{\'a}ez, and Luc~Van Gool.
\newblock Deep retinal image understanding.
\newblock In \emph{International conference on medical image computing and computer-assisted intervention}, pages 140--148. Springer, 2016.

\bibitem[Mendon{\c{c}}a et~al.(2013)Mendon{\c{c}}a, Ferreira, Marques, Marcal, and Rozeira]{ph2}
Teresa Mendon{\c{c}}a, Pedro~M Ferreira, Jorge~S Marques, Andr{\'e}~RS Marcal, and Jorge Rozeira.
\newblock Ph 2-a dermoscopic image database for research and benchmarking.
\newblock In \emph{2013 35th annual international conference of the IEEE engineering in medicine and biology society (EMBC)}, pages 5437--5440. IEEE, 2013.

\bibitem[Meyer et~al.(2017)Meyer, Costa, Galdran, Mendon{\c{c}}a, and Campilho]{mdeep}
Maria~Ines Meyer, Pedro Costa, Adrian Galdran, Ana~Maria Mendon{\c{c}}a, and Aur{\'e}lio Campilho.
\newblock A deep neural network for vessel segmentation of scanning laser ophthalmoscopy images.
\newblock In \emph{International Conference Image Analysis and Recognition}, pages 507--515. Springer, 2017.

\bibitem[Odstrcilik et~al.(2013)Odstrcilik, Kolar, Budai, Hornegger, Jan, Gazarek, Kubena, Cernosek, Svoboda, and Angelopoulou]{hrf}
Jan Odstrcilik, Radim Kolar, Attila Budai, Joachim Hornegger, Jiri Jan, Jiri Gazarek, Tomas Kubena, Pavel Cernosek, Ondrej Svoboda, and Elli Angelopoulou.
\newblock Retinal vessel segmentation by improved matched filtering: evaluation on a new high-resolution fundus image database.
\newblock \emph{IET Image Processing}, 7\penalty0 (4):\penalty0 373--383, 2013.

\bibitem[Oktay et~al.(2018)Oktay, Schlemper, Folgoc, Lee, Heinrich, Misawa, Mori, McDonagh, Hammerla, Kainz, et~al.]{attentionunet}
Ozan Oktay, Jo~Schlemper, Loic~Le Folgoc, Matthew Lee, Mattias Heinrich, Kazunari Misawa, Kensaku Mori, Steven McDonagh, Nils~Y Hammerla, Bernhard Kainz, et~al.
\newblock Attention u-net: Learning where to look for the pancreas.
\newblock \emph{arXiv preprint arXiv:1804.03999}, 2018.

\bibitem[Paszke et~al.(2019)Paszke, Gross, Massa, Lerer, Bradbury, Chanan, Killeen, Lin, Gimelshein, Antiga, et~al.]{pytorch}
Adam Paszke, Sam Gross, Francisco Massa, Adam Lerer, James Bradbury, Gregory Chanan, Trevor Killeen, Zeming Lin, Natalia Gimelshein, Luca Antiga, et~al.
\newblock Pytorch: An imperative style, high-performance deep learning library.
\newblock \emph{Advances in neural information processing systems}, 32, 2019.

\bibitem[Pham et~al.(2000)Pham, Xu, and Prince]{pham2000survey}
Dzung~L Pham, Chenyang Xu, and Jerry~L Prince.
\newblock A survey of current methods in medical image segmentation.
\newblock \emph{Annual review of biomedical engineering}, 2\penalty0 (3):\penalty0 315--337, 2000.

\bibitem[Ronneberger et~al.(2015)Ronneberger, Fischer, and Brox]{unet}
Olaf Ronneberger, Philipp Fischer, and Thomas Brox.
\newblock U-net: Convolutional networks for biomedical image segmentation.
\newblock In \emph{International Conference on Medical image computing and computer-assisted intervention}, pages 234--241. Springer, 2015.

\bibitem[Ruder(2016)]{sgd}
Sebastian Ruder.
\newblock An overview of gradient descent optimization algorithms.
\newblock \emph{arXiv preprint arXiv:1609.04747}, 2016.

\bibitem[Shamshad et~al.(2022)Shamshad, Khan, Zamir, Khan, Hayat, Khan, and Fu]{shamshad2022transformers}
Fahad Shamshad, Salman Khan, Syed~Waqas Zamir, Muhammad~Haris Khan, Munawar Hayat, Fahad~Shahbaz Khan, and Huazhu Fu.
\newblock Transformers in medical imaging: A survey.
\newblock \emph{arXiv preprint arXiv:2201.09873}, 2022.

\bibitem[Sharma and Aggarwal(2010)]{sharma2010automated}
Neeraj Sharma and Lalit~M Aggarwal.
\newblock Automated medical image segmentation techniques.
\newblock \emph{Journal of medical physics/Association of Medical Physicists of India}, 35\penalty0 (1):\penalty0 3, 2010.

\bibitem[Silva et~al.(2014)Silva, Histace, Romain, Dray, and Granado]{etis}
Juan Silva, Aymeric Histace, Olivier Romain, Xavier Dray, and Bertrand Granado.
\newblock Toward embedded detection of polyps in wce images for early diagnosis of colorectal cancer.
\newblock \emph{International journal of computer assisted radiology and surgery}, 9\penalty0 (2):\penalty0 283--293, 2014.

\bibitem[Tajbakhsh et~al.(2015)Tajbakhsh, Gurudu, and Liang]{colondb}
Nima Tajbakhsh, Suryakanth~R Gurudu, and Jianming Liang.
\newblock Automated polyp detection in colonoscopy videos using shape and context information.
\newblock \emph{IEEE transactions on medical imaging}, 35\penalty0 (2):\penalty0 630--644, 2015.

\bibitem[Tang et~al.(2021)Tang, Li, Zhong, Ding, and Song]{tang2021disentangled}
Lv~Tang, Bo~Li, Yijie Zhong, Shouhong Ding, and Mofei Song.
\newblock Disentangled high quality salient object detection.
\newblock In \emph{Proceedings of the IEEE/CVF International Conference on Computer Vision}, pages 3580--3590, 2021.

\bibitem[Vaswani et~al.(2017)Vaswani, Shazeer, Parmar, Uszkoreit, Jones, Gomez, Kaiser, and Polosukhin]{vaswani2017attention}
Ashish Vaswani, Noam Shazeer, Niki Parmar, Jakob Uszkoreit, Llion Jones, Aidan~N Gomez, {\L}ukasz Kaiser, and Illia Polosukhin.
\newblock Attention is all you need.
\newblock \emph{Advances in neural information processing systems}, 30, 2017.

\bibitem[V{\'a}zquez et~al.(2017)V{\'a}zquez, Bernal, S{\'a}nchez, Fern{\'a}ndez-Esparrach, L{\'o}pez, Romero, Drozdzal, and Courville]{endoscene}
David V{\'a}zquez, Jorge Bernal, F~Javier S{\'a}nchez, Gloria Fern{\'a}ndez-Esparrach, Antonio~M L{\'o}pez, Adriana Romero, Michal Drozdzal, and Aaron Courville.
\newblock A benchmark for endoluminal scene segmentation of colonoscopy images.
\newblock \emph{Journal of healthcare engineering}, 2017, 2017.

\bibitem[Wang et~al.(2021)Wang, Wei, Wang, Zhou, Zhu, and Qin]{wang2021boundary}
Jiacheng Wang, Lan Wei, Liansheng Wang, Qichao Zhou, Lei Zhu, and Jing Qin.
\newblock Boundary-aware transformers for skin lesion segmentation.
\newblock In \emph{International Conference on Medical Image Computing and Computer-Assisted Intervention}, pages 206--216. Springer, 2021.

\bibitem[Wang et~al.(1998)Wang, Adali, Kung, and Szabo]{wang1998quantification}
Yue Wang, T{\"u}lay Adali, Sun-Yuan Kung, and Zsolt Szabo.
\newblock Quantification and segmentation of brain tissues from mr images: A probabilistic neural network approach.
\newblock \emph{IEEE transactions on image processing}, 7\penalty0 (8):\penalty0 1165--1181, 1998.

\bibitem[Wei et~al.(2021)Wei, Hu, Zhang, Li, Zhou, and Cui]{sarnet}
Jun Wei, Yiwen Hu, Ruimao Zhang, Zhen Li, S~Kevin Zhou, and Shuguang Cui.
\newblock Shallow attention network for polyp segmentation.
\newblock In \emph{International Conference on Medical Image Computing and Computer-Assisted Intervention}, pages 699--708. Springer, 2021.

\bibitem[Xie et~al.(2022)Xie, Xia, Ma, Zhao, Chen, and Li]{xie2022pyramid}
Chenxi Xie, Changqun Xia, Mingcan Ma, Zhirui Zhao, Xiaowu Chen, and Jia Li.
\newblock Pyramid grafting network for one-stage high resolution saliency detection.
\newblock In \emph{Proceedings of the IEEE/CVF Conference on Computer Vision and Pattern Recognition}, pages 11717--11726, 2022.

\bibitem[Xie and Tu(2015)]{hed}
Saining Xie and Zhuowen Tu.
\newblock Holistically-nested edge detection.
\newblock In \emph{Proceedings of the IEEE international conference on computer vision}, pages 1395--1403, 2015.

\bibitem[Zeng et~al.(2019)Zeng, Zhang, Zhang, Lin, and Lu]{zeng2019towards}
Yi~Zeng, Pingping Zhang, Jianming Zhang, Zhe Lin, and Huchuan Lu.
\newblock Towards high-resolution salient object detection.
\newblock In \emph{Proceedings of the IEEE/CVF International Conference on Computer Vision}, pages 7234--7243, 2019.

\bibitem[{Zhang} et~al.(2016){Zhang}, {Dashtbozorg}, {Bekkers}, {Pluim}, {Duits}, and {ter Haar Romeny}]{iostar}
J.~{Zhang}, B.~{Dashtbozorg}, E.~{Bekkers}, J.~P.~W. {Pluim}, R.~{Duits}, and B.~M. {ter Haar Romeny}.
\newblock Robust retinal vessel segmentation via locally adaptive derivative frames in orientation scores.
\newblock \emph{IEEE Transactions on Medical Imaging}, 35\penalty0 (12):\penalty0 2631--2644, Dec 2016.
\newblock ISSN 0278-0062.

\bibitem[Zhang et~al.(2021)Zhang, Liu, and Hu]{zhang2021transfuse}
Yundong Zhang, Huiye Liu, and Qiang Hu.
\newblock Transfuse: Fusing transformers and cnns for medical image segmentation.
\newblock In \emph{International Conference on Medical Image Computing and Computer-Assisted Intervention}, pages 14--24. Springer, 2021.

\bibitem[Zhou et~al.(2018)Zhou, Rahman~Siddiquee, Tajbakhsh, and Liang]{unet++}
Zongwei Zhou, Md~Mahfuzur Rahman~Siddiquee, Nima Tajbakhsh, and Jianming Liang.
\newblock Unet++: A nested u-net architecture for medical image segmentation.
\newblock In \emph{Deep learning in medical image analysis and multimodal learning for clinical decision support}, pages 3--11. Springer, 2018.

\end{thebibliography}
\newpage
\appendix 
\section*{Appendix}\vspace{-0.5em}
In this Appendix document, we have provided information about the dataset we used, additional results from the experiments, that we did not mention in paper due to page limitation constraints. We believe this supplementary information will help the scientific community to understand and reproduce our conducted research in a better way.\vspace{-1em}

\section{Datasets}\label{sec:datasets}\vspace{-0.5em}
We have used ten datasets for three different segmentation tasks: \textbf{(a)} skin lesion segmentation (2 datasets),  \textbf{(b)} retinal vessel segmentation (3 datasets), and \textbf{(c)} polyp segmentation (5 datasets). An overview of all these publicly available dataset in mentioned in the Table \ref{table:datasets}.

\begin{table}[!h]
\centering
    \begin{minipage}{.96\textwidth}
        \centering\small
        \setlength{\tabcolsep}{4pt}

        \scalebox{0.8}[0.8]{
\begin{tabular}{lccc}
%column{3} = {teal7},
%cell{2}{3} = {yellow7},
\hline
\hline
\rowcolor{gray}
\textbf{Dataset} & \textbf{Average Resolution} &{\textbf{Train Samples}}&{\textbf{Test Samples}}\\
\hline
\underline{\textbf{Skin lesion}}&&&\\
ISIC-2016 & 1468 $\times$ 1070 & 900 & 379\\
PH2 & 766 $\times$ 575 & 0 & 200 \\
\hline
\underline{\textbf{Retinal Vessel}}&&&\\
HRF& 3504 $\times$ 2306 & 30& 15\\
IOSTAR& 1024 $\times$ 1024 &20 &10\\
CHASE& 999 $\times$ 960 &20 &8\\
\hline
\underline{\textbf{Polyp}}&&&\\
Kvasir&618 $\times$ 539&838&100\\
ClinicDB&384 $\times$ 288&612&62\\
ColonDB&574 $\times$ 500&0&380\\
Endoscene&574 $\times$ 500&0&60\\
ETIS&1225 $\times$ 966&0&196\\
\hline
\end{tabular}}
\end{minipage}
\vfill
   \begin{minipage}{.96\textwidth}
\centering
\caption{\footnotesize{An overview of dataset used in the paper for three segmentation tasks.}}%
\label{table:datasets}
    \end{minipage} 
    \vspace{-2em}
\end{table}

\section{Probability Correction Strategy (PCS)}\label{sec:PCS}
\begin{figure*}[!ht]
\begin{minipage}{0.5\textwidth}
\centering
  \begin{minipage}{0.46\textwidth}
      \centering
       % lelf lower right up trim= 7.5mm 0mm 0mm 10mm
   \includegraphics[height=2.4cm, width=\linewidth]{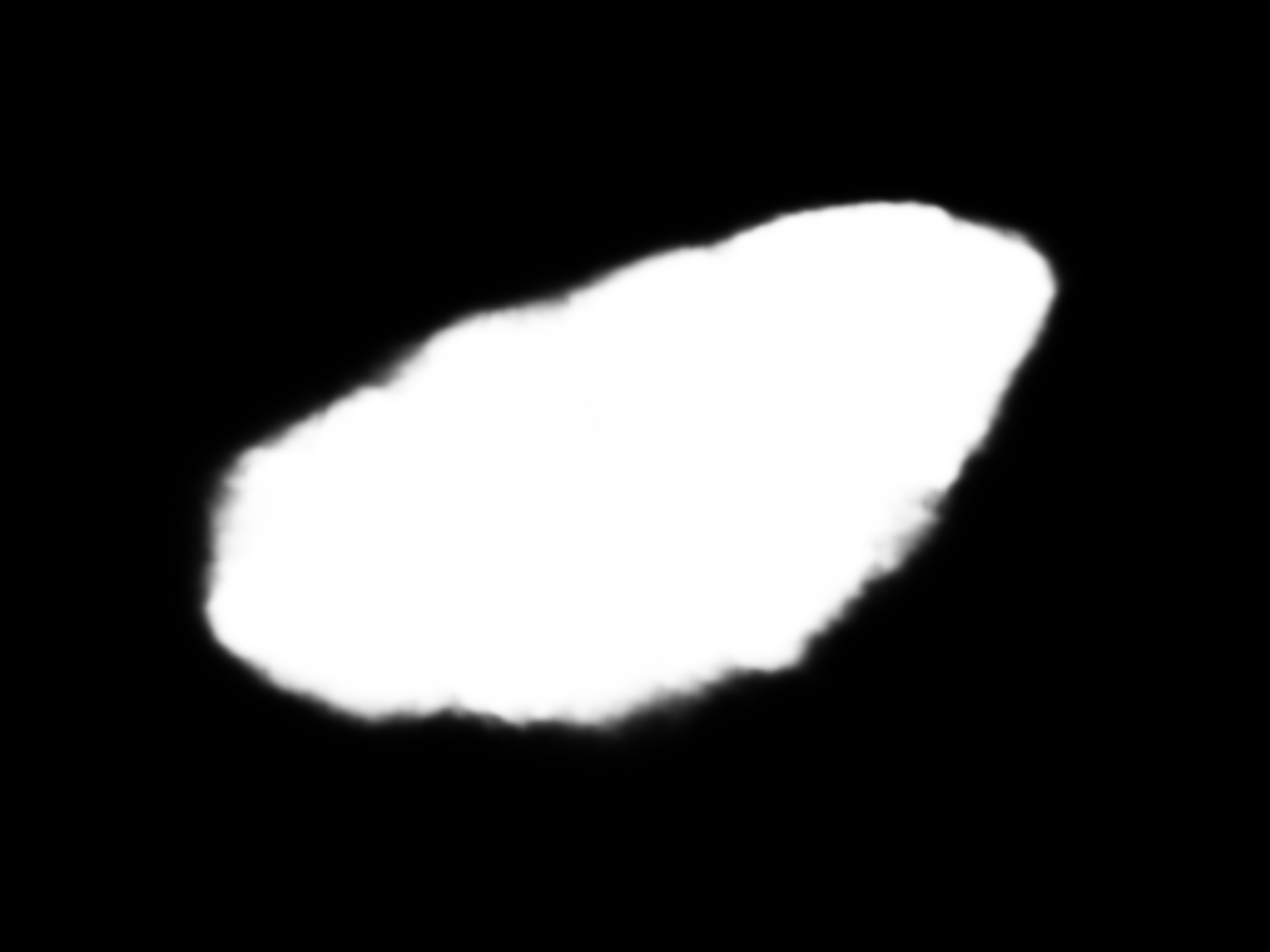}
   \tiny {\textit{IoU}= 0.835}  
   \vspace{0.3em}
  \end{minipage}
\begin{minipage}{0.46\textwidth}
      \centering
       % lelf lower right up
  \includegraphics[height=2.4cm, width=\linewidth]{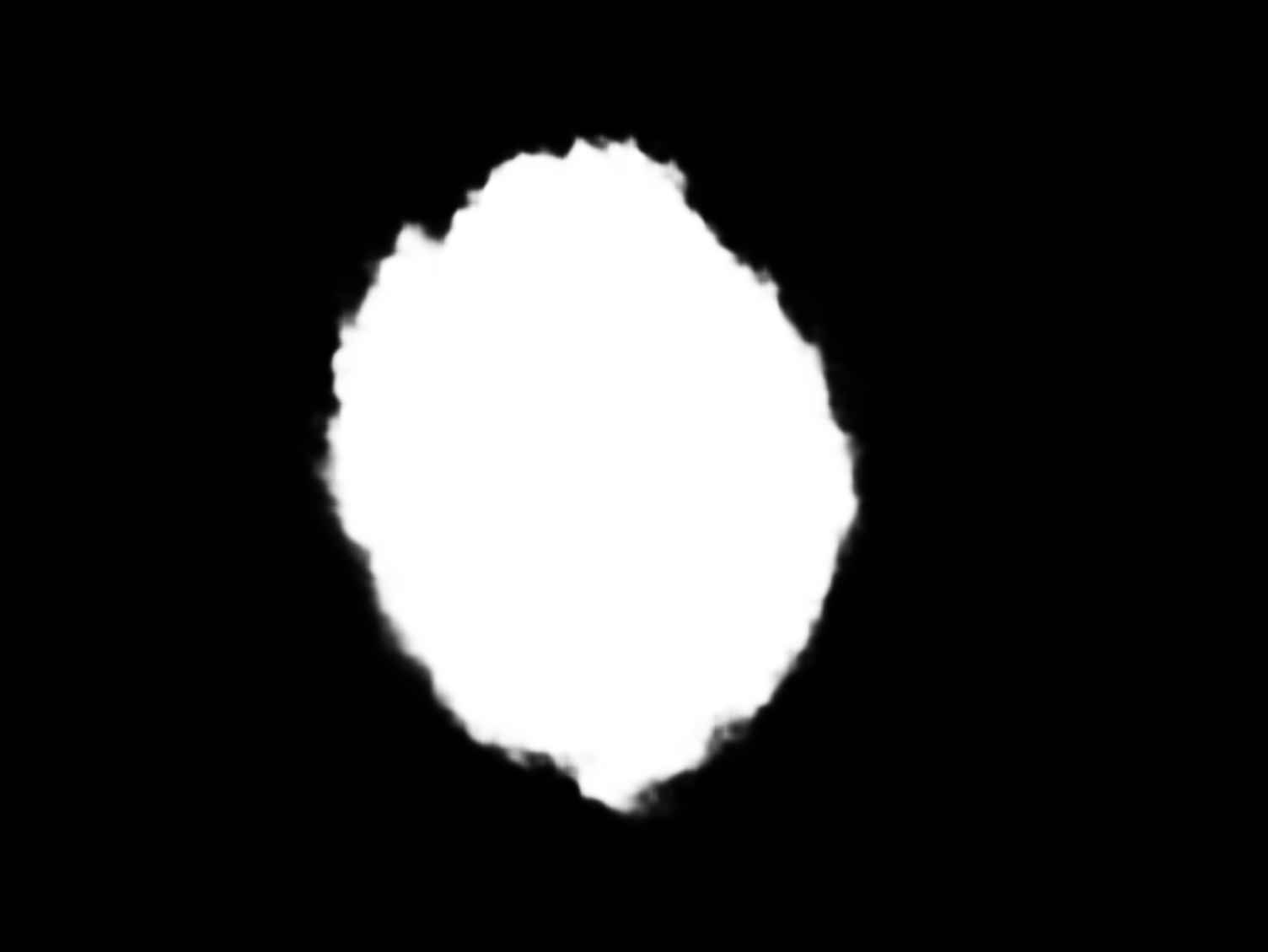}
\tiny {\textit{IoU}= 0.892}  
   \vspace{0.3em}
 %\vspace{0.1em}
  \end{minipage}
  \footnotesize Predicted Mask without PCS
  \end{minipage}
  \hfill
\begin{minipage}{0.5\textwidth}
\centering
\begin{minipage}{0.46\textwidth}
      \centering
       % lelf lower right up
  \includegraphics[height=2.4cm, width=\linewidth]{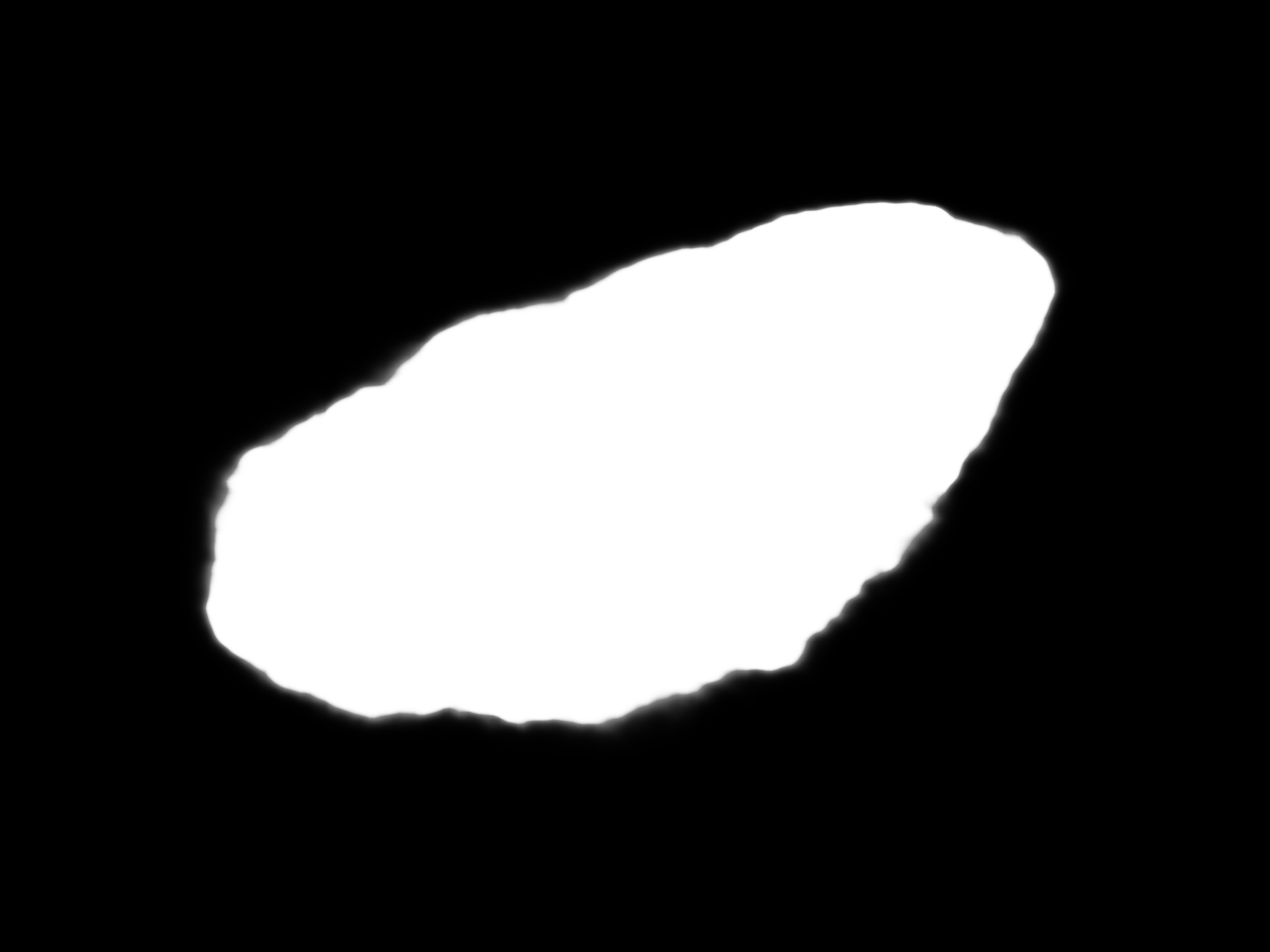}
\tiny {\textit{IoU}= 0.839}  
   \vspace{0.3em}
  \end{minipage}
  \begin{minipage}{0.46\textwidth}
      \centering
       % lelf lower right up trim= 7.5mm 0mm 0mm 10mm
    \includegraphics[height=2.4cm, width=\linewidth]{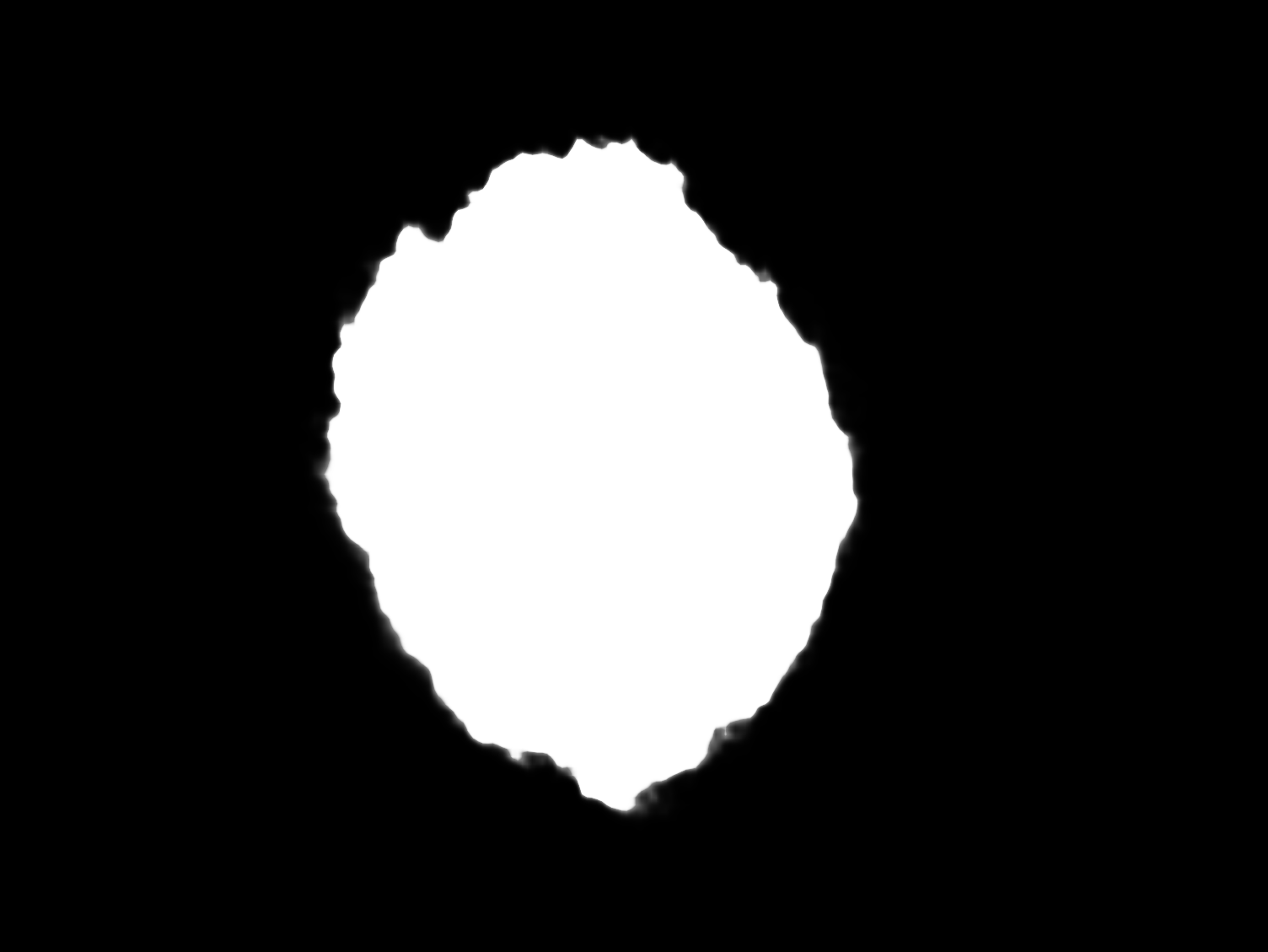}
\tiny {\textit{IoU}= 0.902}  
   \vspace{0.3em}
  \end{minipage}
  \footnotesize Predicted Mask with PCS
  \end{minipage}
  \vfill
  \begin{minipage}{0.96\textwidth}
\caption{ \footnotesize{Visualizations of predicted mask with PCS. The left side two images shows the predicted mask without applying PCS while right side represents the sharp mask after applying PCS. The IoU scores represents that performance increases after applying PCS.}}\label{fig:visualization}
  \end{minipage}
\end{figure*}

We have applied Probability Correction Strategy (PCS) \cite{sarnet} during inference to improve the final prediction. During the training pipeline, the number of negative samples (background pixels) is greater than the number of positive samples (foreground pixels), which leads the model to produce an unsharp and noisy output. We enhance the final output through logits re-weighting. In PCS, we count the number of samples of each class (positive and negative pixels) before sigmoid function and normalize the logits of each class with the total count of corresponding class. Figure \ref{fig:visualization}, demonstrates the visualizations of predicted mask before and after applying PCS.

\section{Additional Qualitative Results}\label{sec:add-visual-results}
\begin{figure*}[!ht]
\begin{minipage}{0.90\textwidth}
\hspace{2.5em}
\centering
  \begin{minipage}{0.30\textwidth}
       % lelf lower right up trim= 7.5mm 0mm 0mm 10mm
   \includegraphics[height=3.0cm, width=\linewidth]{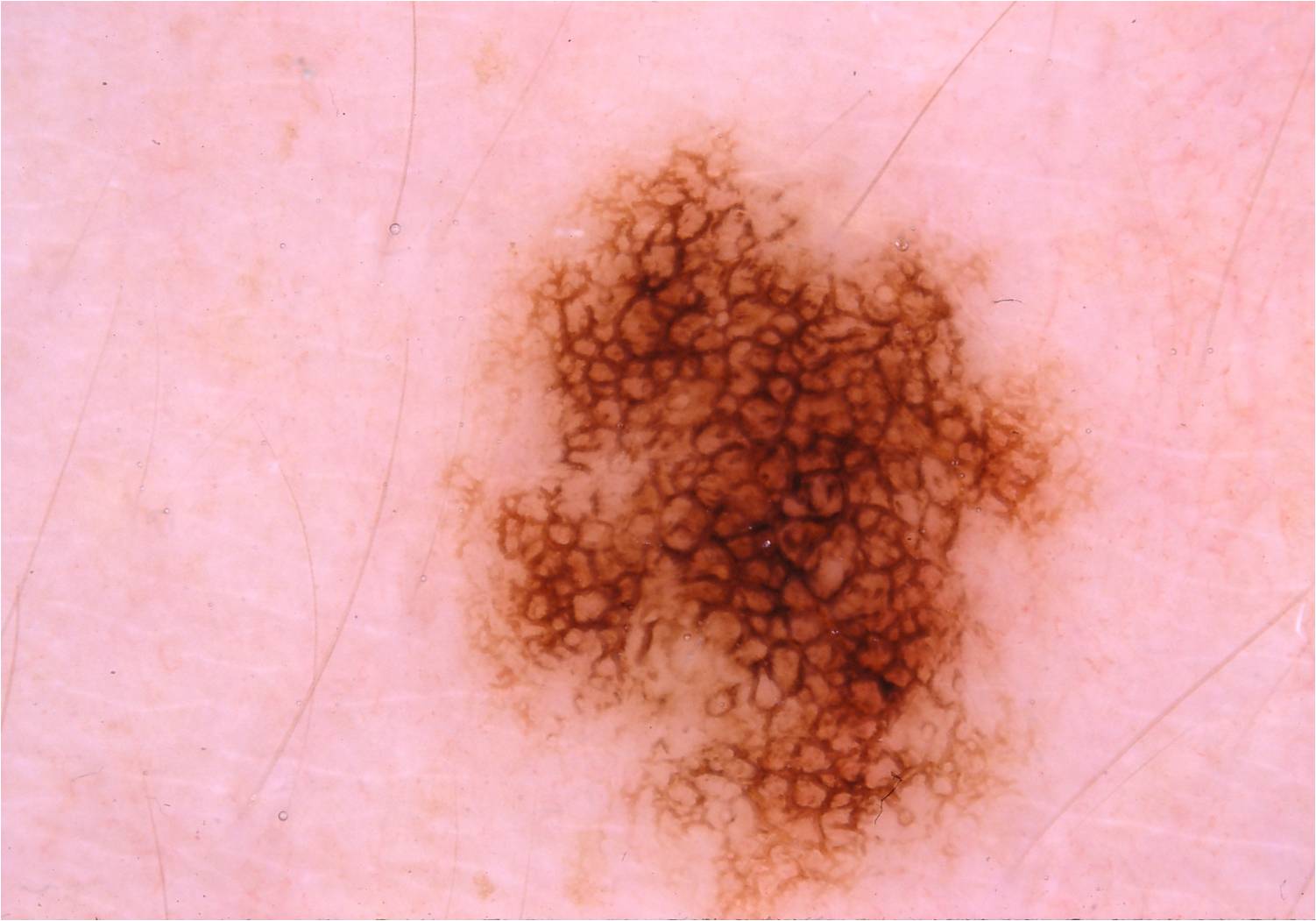}
  \end{minipage}
\begin{minipage}{0.30\textwidth}
\centering
       % lelf lower right up
  \includegraphics[height=3.0cm, width=\linewidth]{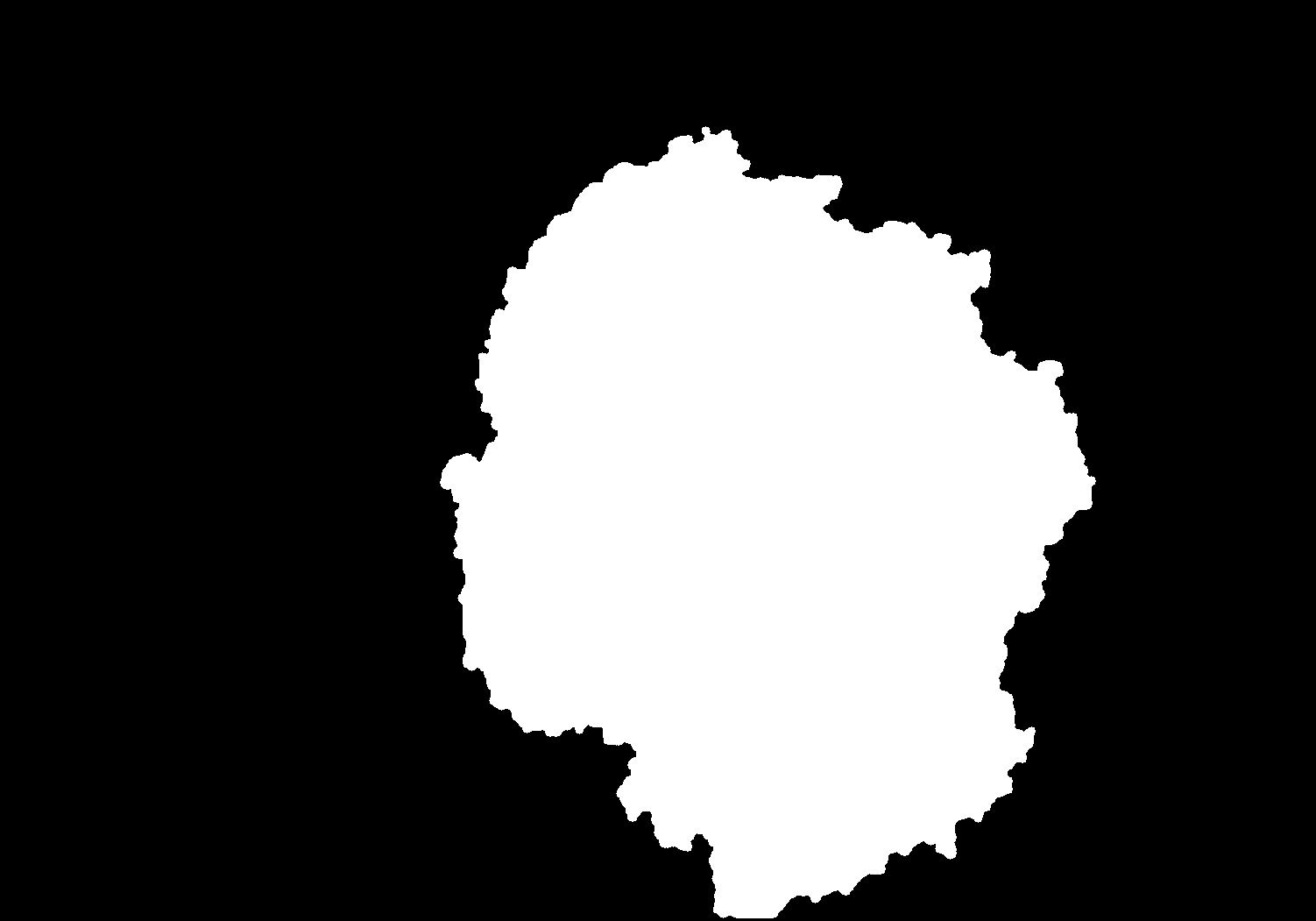}
  \end{minipage}
  \begin{minipage}{0.30\textwidth}
  \centering
       % lelf lower right up
  \includegraphics[height=3.0cm, width=\linewidth]{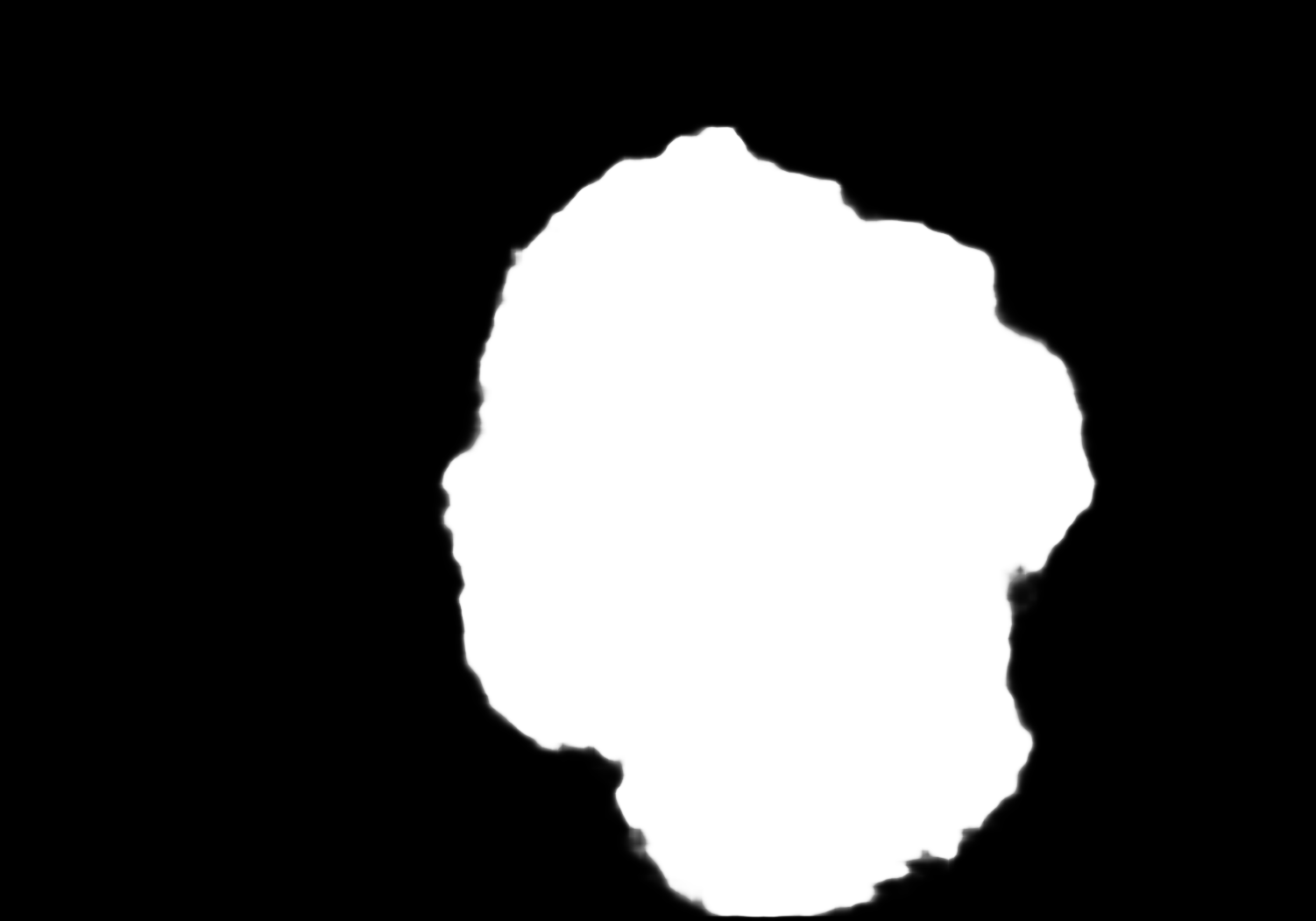}

  \end{minipage}
  \end{minipage}
  \vfill
  
\begin{minipage}{0.90\textwidth}
\centering
  \hspace{2.5em}
  \begin{minipage}{0.30\textwidth}
       % lelf lower right up trim= 7.5mm 0mm 0mm 10mm
   \includegraphics[height=3.0cm, width=\linewidth]{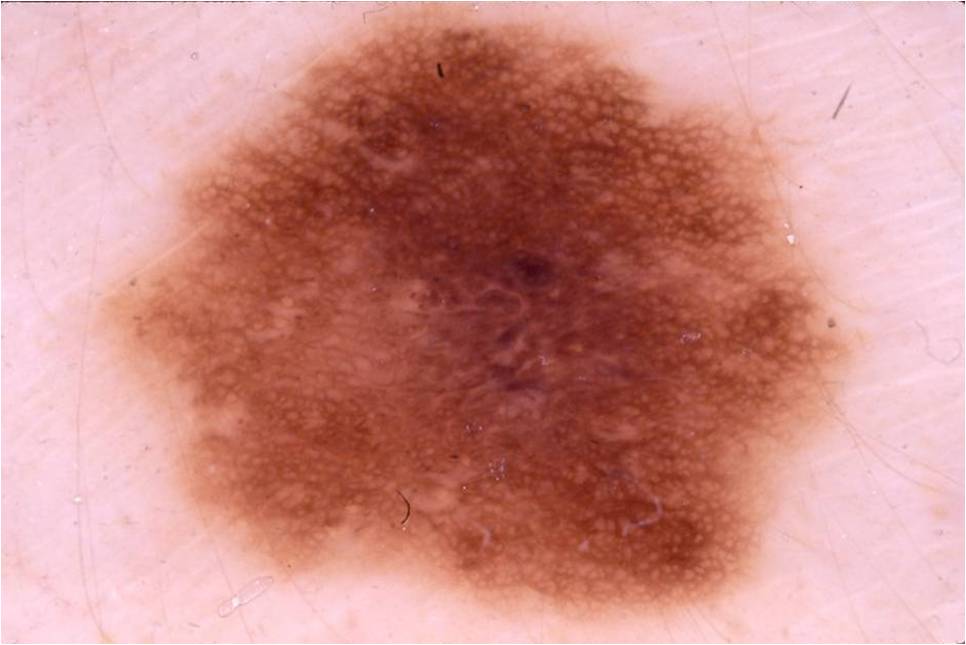}
  \end{minipage}
\begin{minipage}{0.30\textwidth}
      \centering
       % lelf lower right up
  \includegraphics[height=3.0cm, width=\linewidth]{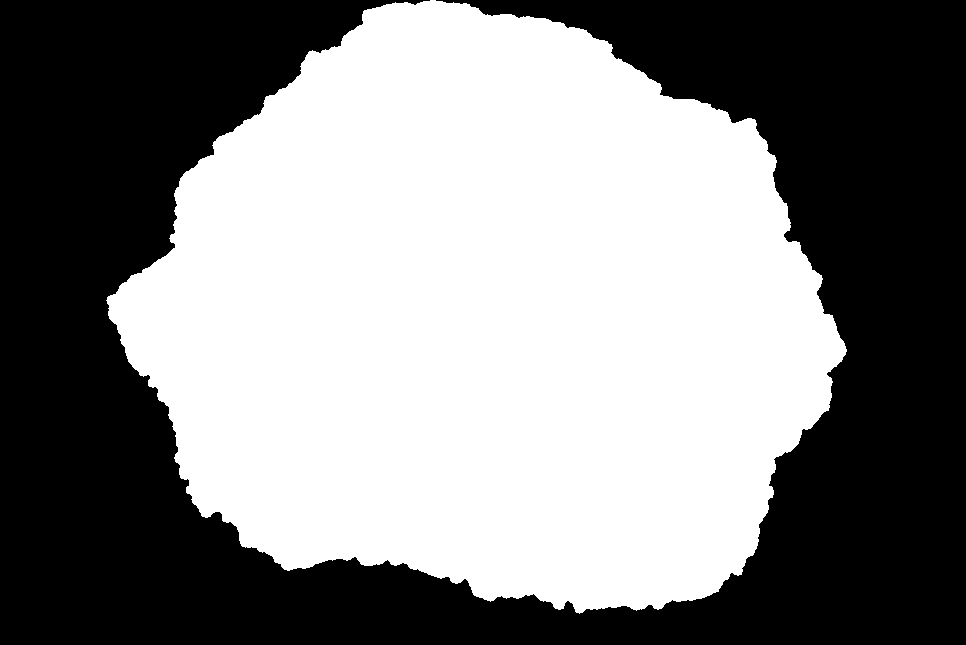}
  \end{minipage}
  \begin{minipage}{0.30\textwidth}
      \centering
       % lelf lower right up
  \includegraphics[height=3.0cm, width=\linewidth]{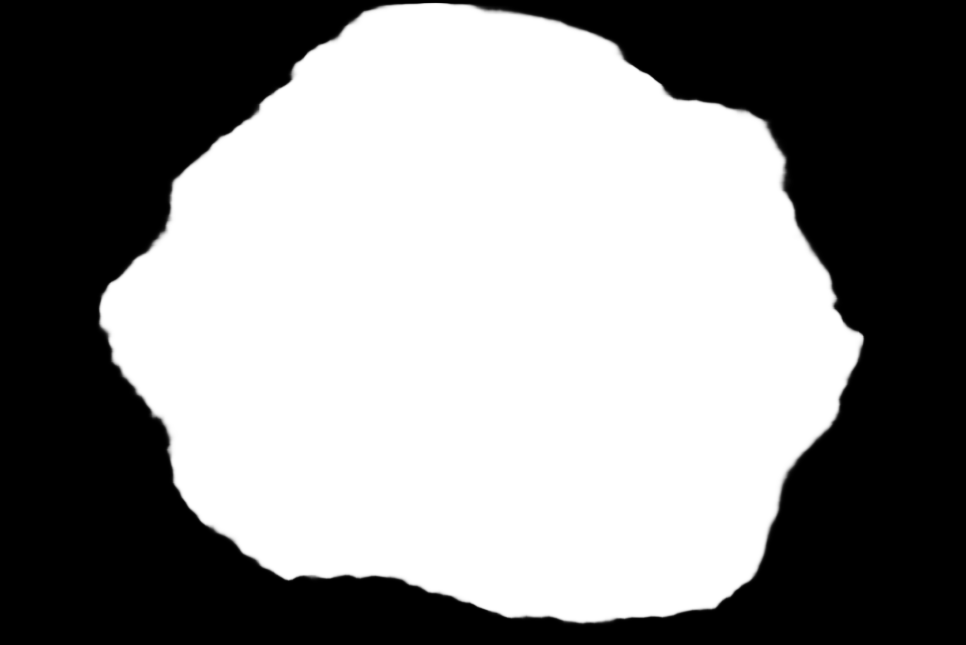}

  \end{minipage}
  \end{minipage}
  \vfill
\begin{minipage}{0.90\textwidth}
\centering
  \hspace{2.5em}
  \begin{minipage}{0.30\textwidth}

       % lelf lower right up trim= 7.5mm 0mm 0mm 10mm
   \includegraphics[height=3.0cm, width=\linewidth]{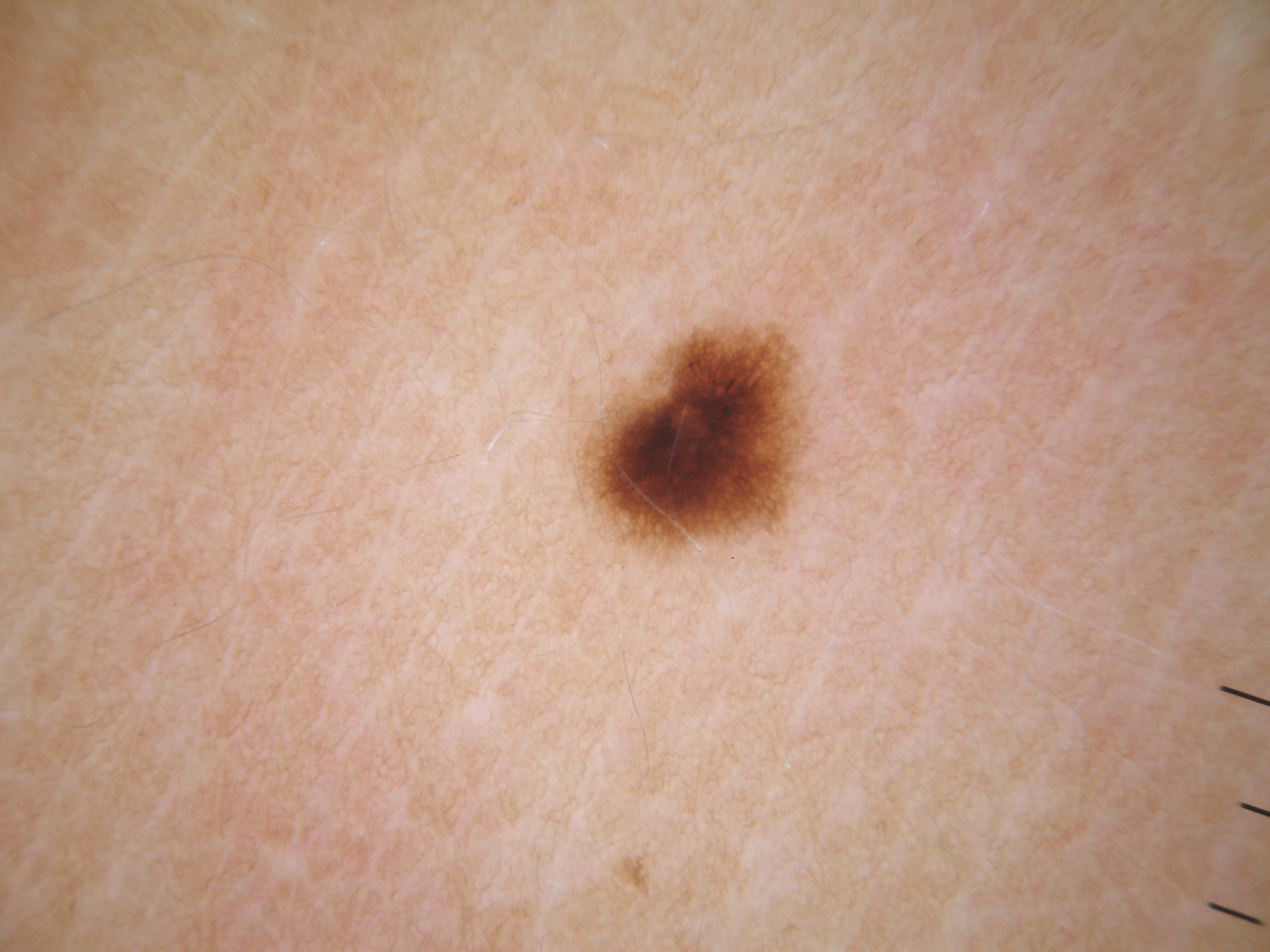}
  \end{minipage}
\begin{minipage}{0.30\textwidth}
      \centering
       % lelf lower right up
  \includegraphics[height=3.0cm, width=\linewidth]{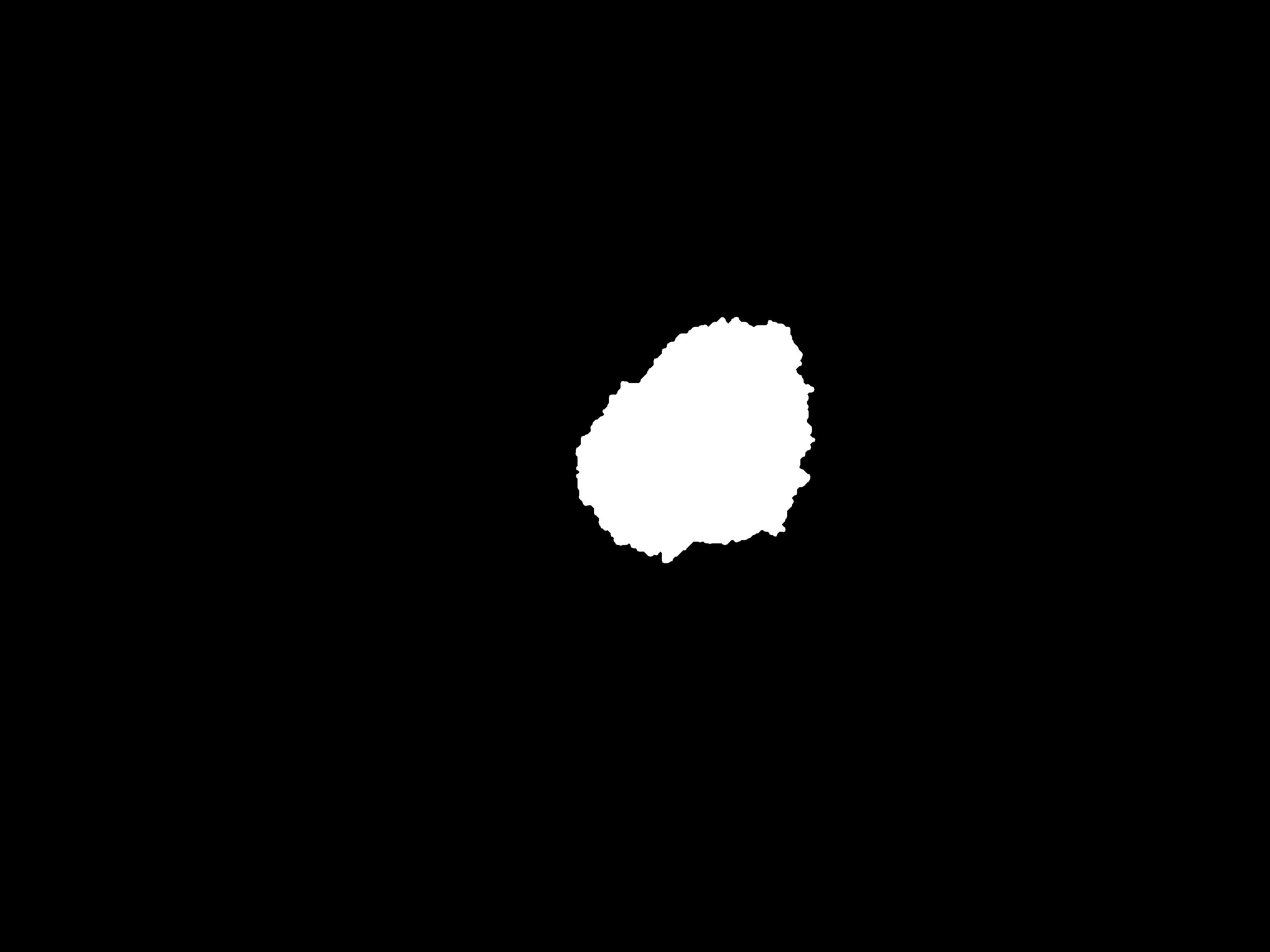}
  \end{minipage}
  \begin{minipage}{0.30\textwidth}
      \centering
       % lelf lower right up
  \includegraphics[height=3.0cm, width=\linewidth]{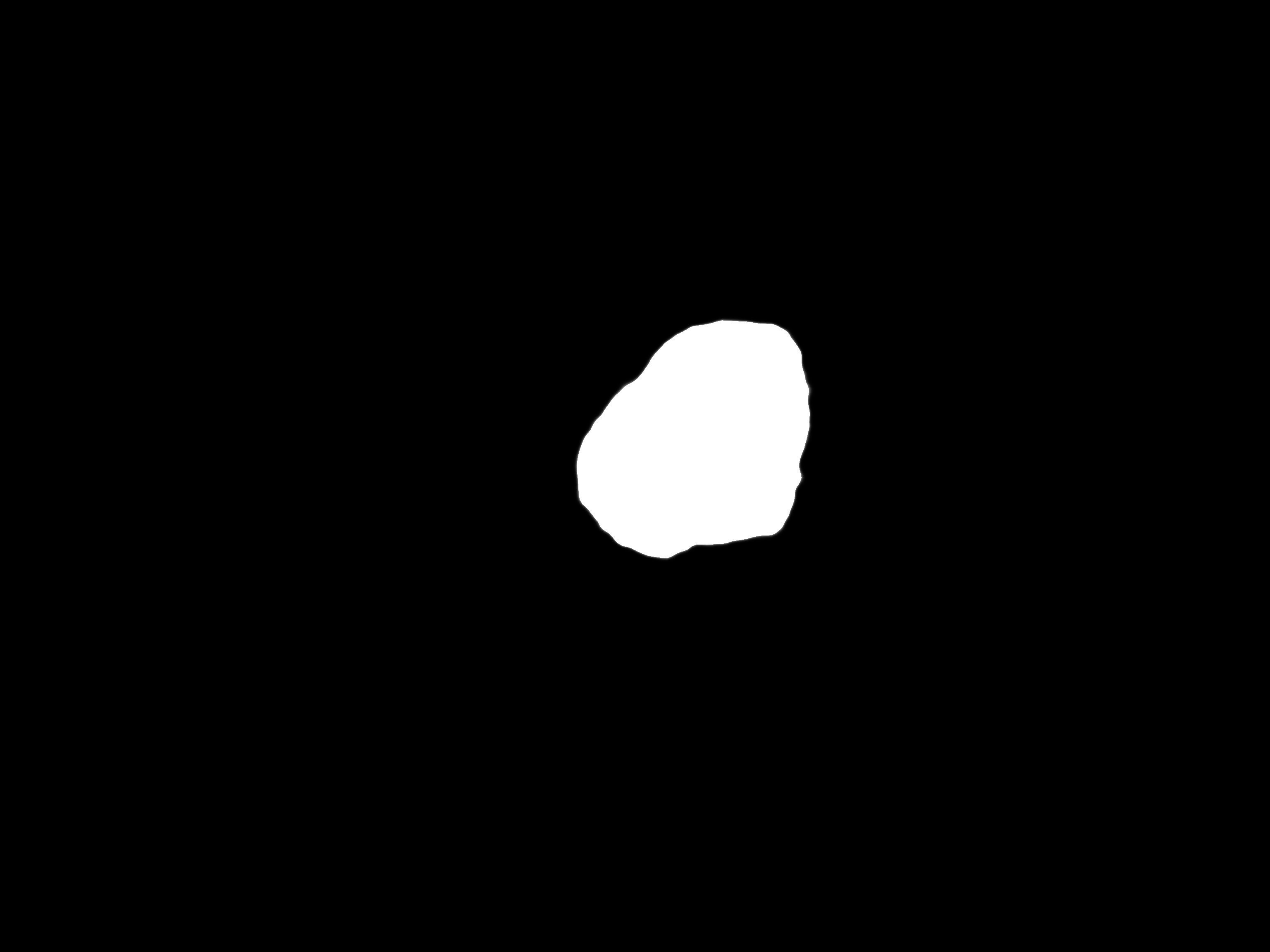}

  \end{minipage}
  \end{minipage}

  \vfill
  
\begin{minipage}{0.90\textwidth}
\centering
  \hspace{2.5em}
  \begin{minipage}{0.30\textwidth}
       % lelf lower right up trim= 7.5mm 0mm 0mm 10mm
   \includegraphics[height=3.0cm, width=\linewidth]{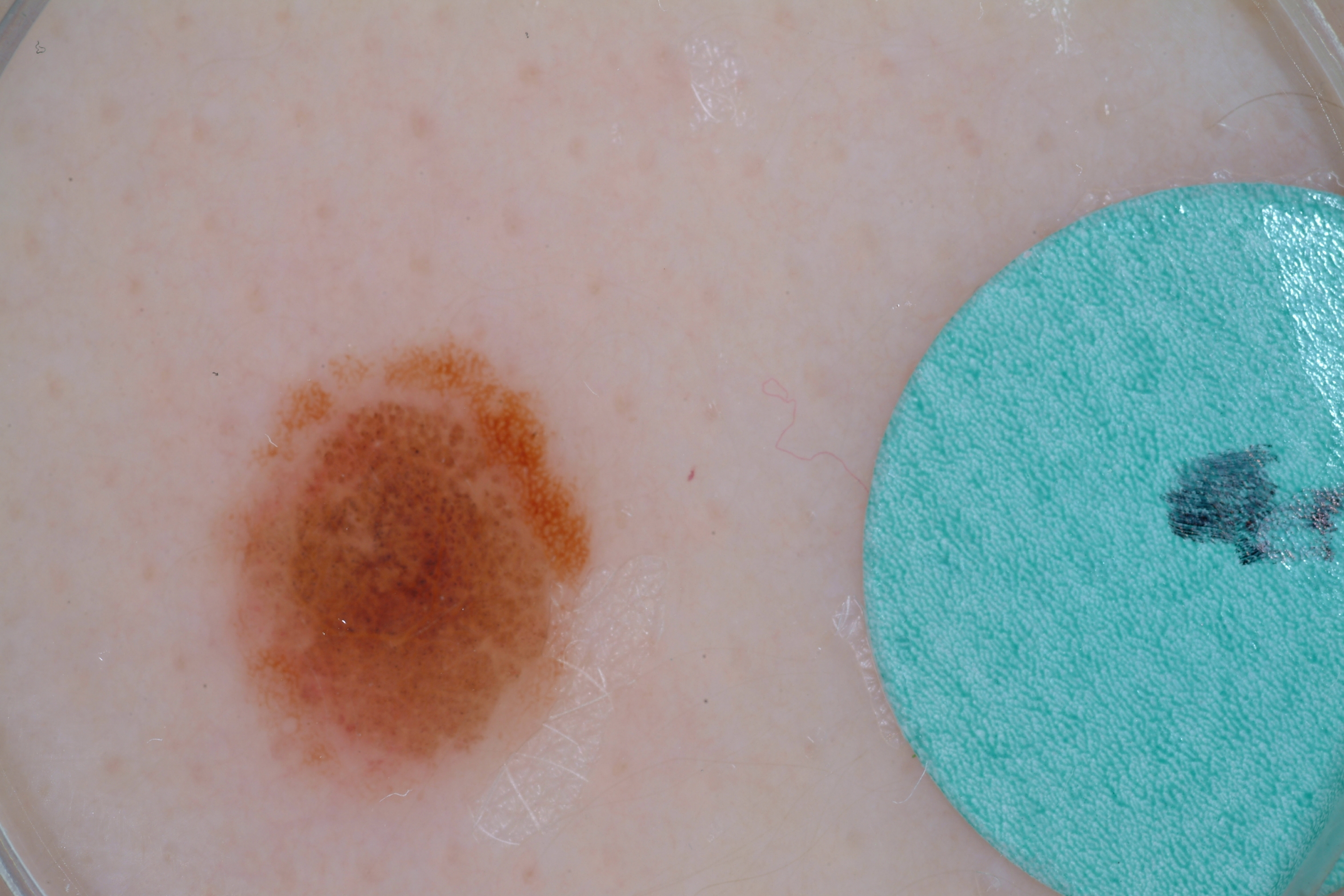}
     \centering \footnotesize \textbf{IMAGE}
  \end{minipage}
\begin{minipage}{0.30\textwidth}
      \centering
       % lelf lower right up
  \includegraphics[height=3.0cm, width=\linewidth]{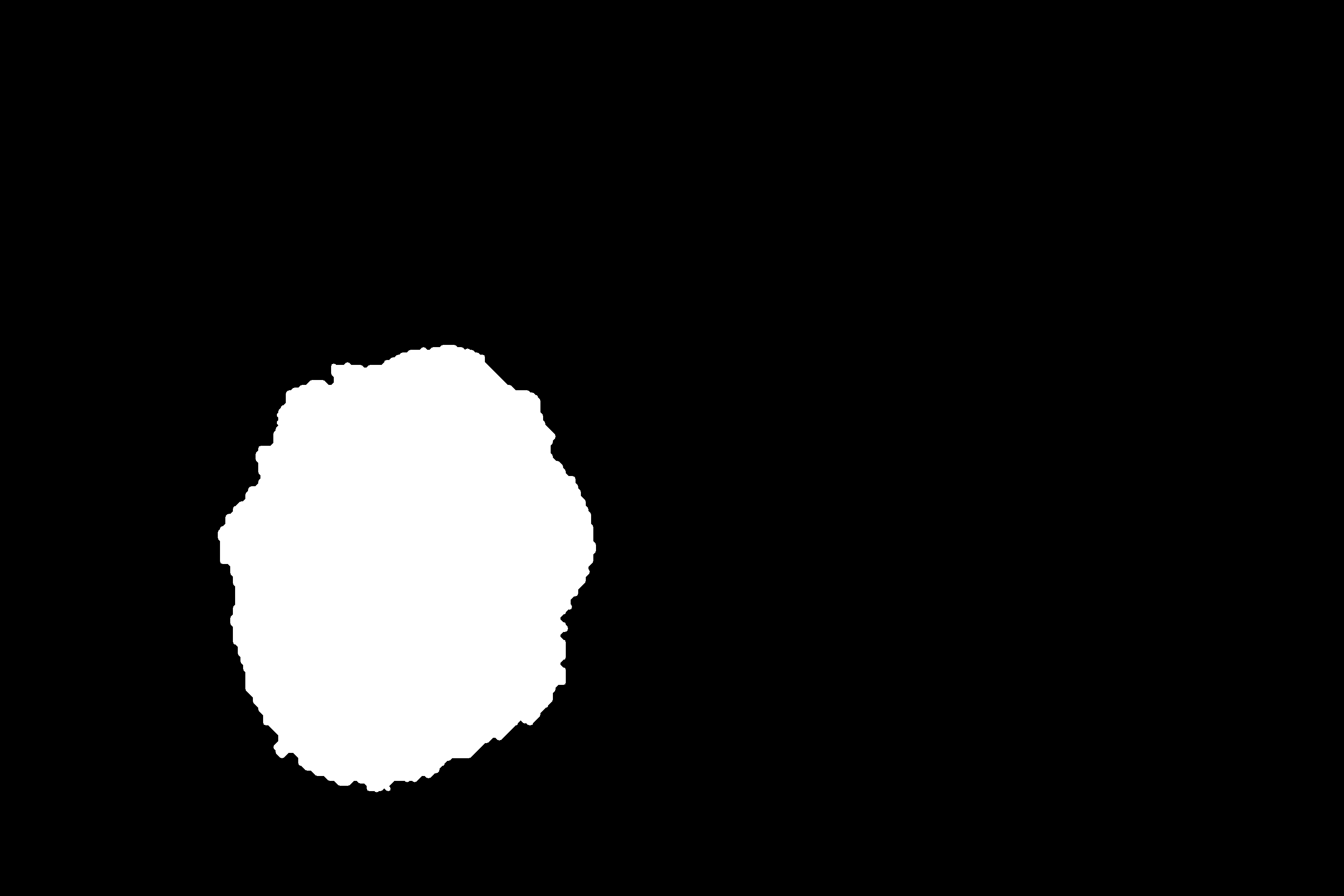}
    \footnotesize \textbf{GROUND TRUTH}
  \end{minipage}
  \begin{minipage}{0.30\textwidth}
      \centering
       % lelf lower right up
  \includegraphics[height=3.0cm, width=\linewidth]{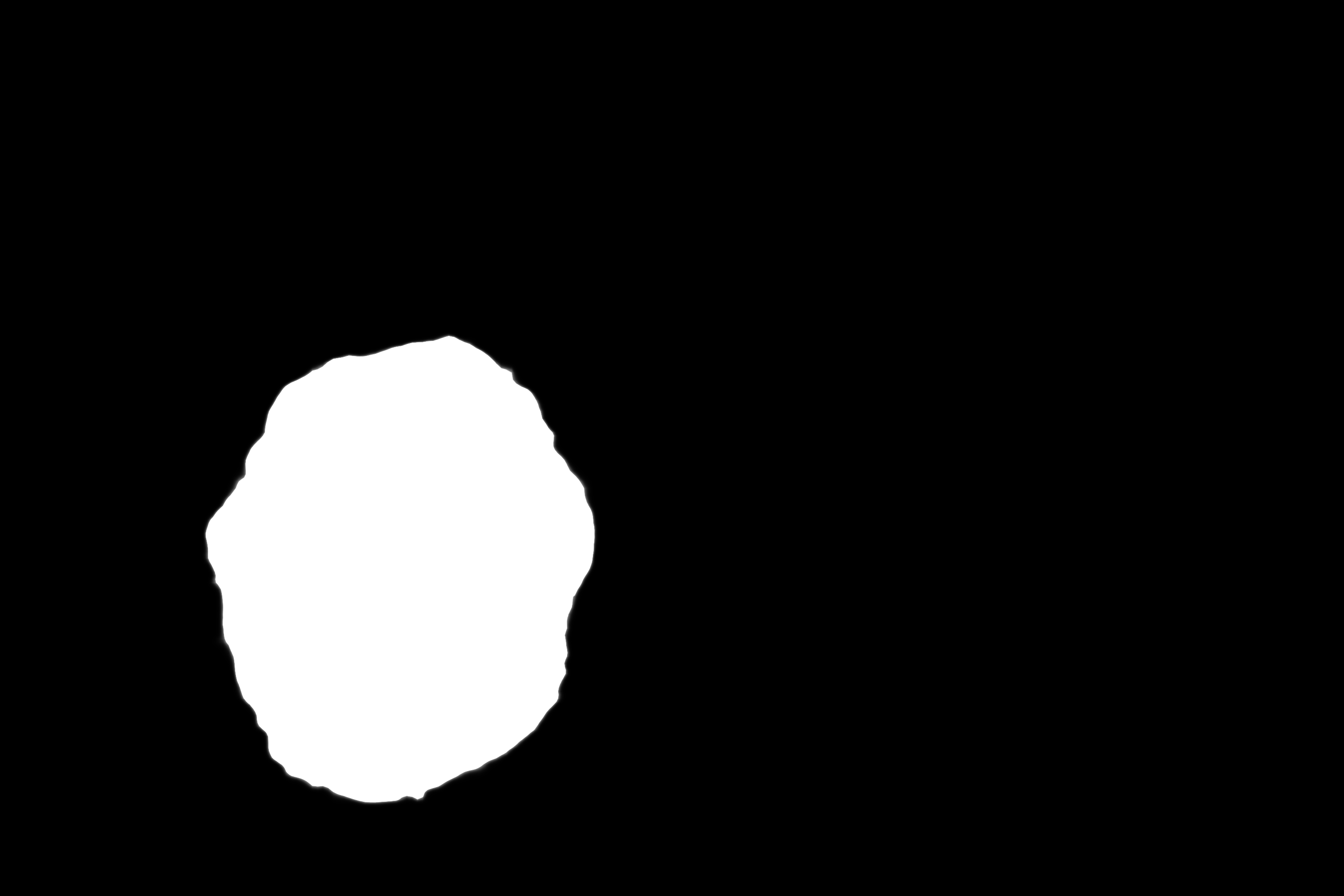}
    \footnotesize \textbf{PREDICTED MASK}

  \end{minipage}
  
    \vfill
  \begin{minipage}{0.96\textwidth}
  \centering
\caption{ \footnotesize{Visualizations of skin lesion segmentation}}\label{fig:skin-results}
  \end{minipage}
  
  \end{minipage}
  %\vspace{-1em}
\end{figure*}

\begin{figure*}[!ht]
\begin{minipage}{0.90\textwidth}
\centering
  \hspace{2.5em}
  \begin{minipage}{0.30\textwidth}
       % lelf lower right up trim= 7.5mm 0mm 0mm 10mm
   \includegraphics[height=3.0cm, width=\linewidth]{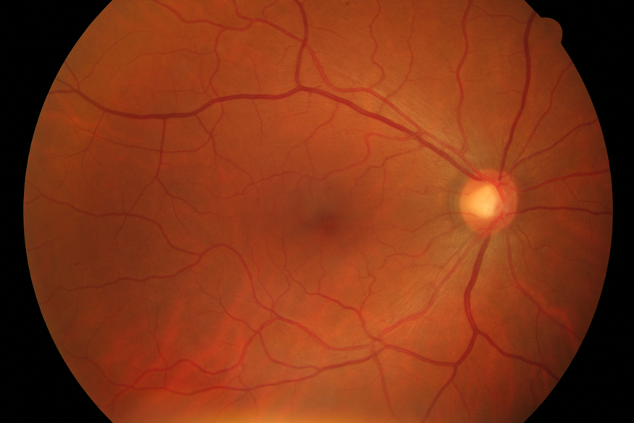}
  \end{minipage}
\begin{minipage}{0.30\textwidth}
      \centering
       % lelf lower right up
  \includegraphics[height=3.0cm, width=\linewidth]{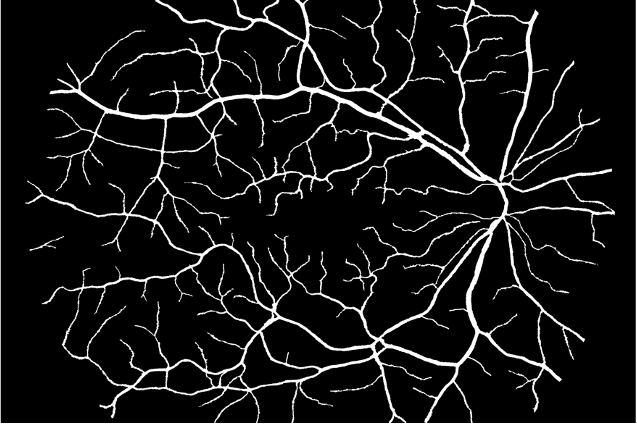}
  \end{minipage}
  \begin{minipage}{0.30\textwidth}
      \centering
       % lelf lower right up
  \includegraphics[height=3.0cm, width=\linewidth]{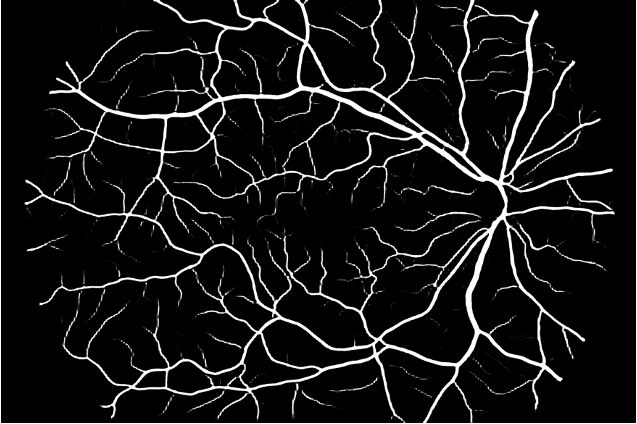}

  \end{minipage}
  \end{minipage}

\begin{minipage}{0.90\textwidth}
\centering
  \hspace{2.5em}
  \begin{minipage}{0.30\textwidth}
       % lelf lower right up trim= 7.5mm 0mm 0mm 10mm
   \includegraphics[height=3.0cm, width=\linewidth]{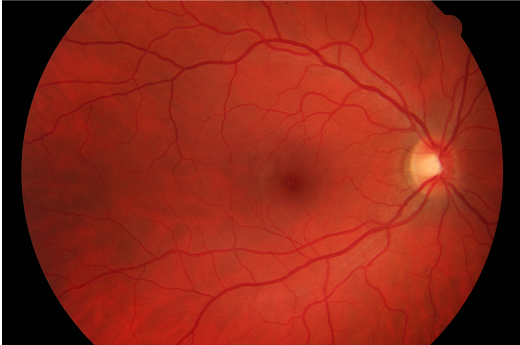}
  \end{minipage}
\begin{minipage}{0.30\textwidth}
      \centering
       % lelf lower right up
  \includegraphics[height=3.0cm, width=\linewidth]{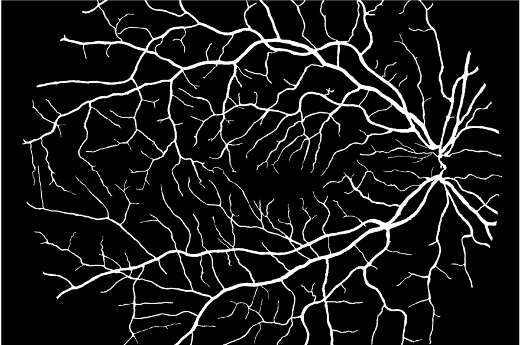}
  \end{minipage}
  \begin{minipage}{0.30\textwidth}
      \centering
       % lelf lower right up
  \includegraphics[height=3.0cm, width=\linewidth]{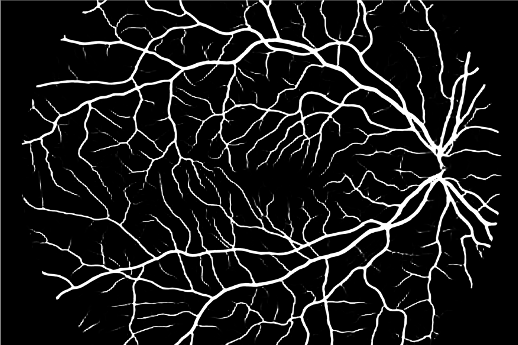}

  \end{minipage}
  \end{minipage}

\vfill
\begin{minipage}{0.90\textwidth}
\centering
  \hspace{2.5em}
  \begin{minipage}{0.30\textwidth}
       % lelf lower right up trim= 7.5mm 0mm 0mm 10mm
   \includegraphics[height=3.0cm, width=\linewidth, angle=180]{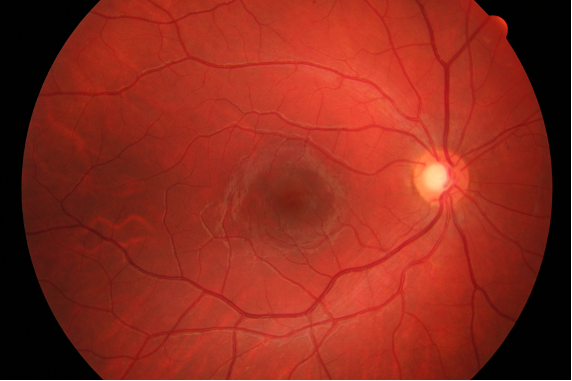}
  \end{minipage}
\begin{minipage}{0.30\textwidth}
      \centering
       % lelf lower right up
  \includegraphics[height=3.0cm, width=\linewidth, angle=180]{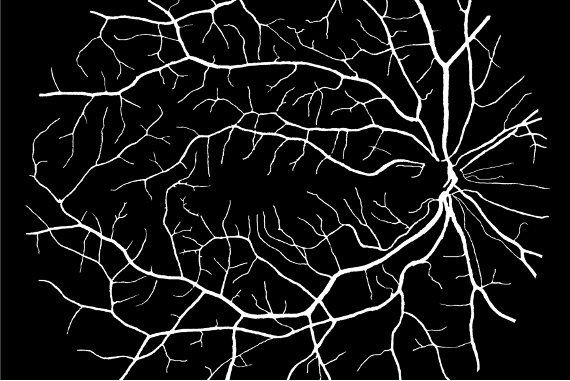}
  \end{minipage}
  \begin{minipage}{0.30\textwidth}
      \centering
       % lelf lower right up
  \includegraphics[height=3.0cm, width=\linewidth, angle=180]{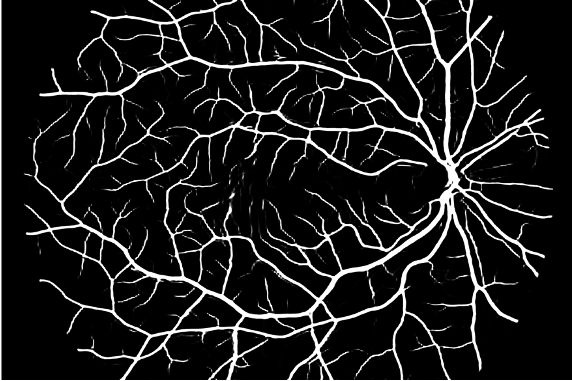}
  \end{minipage}
  \end{minipage}
\vfill
\begin{minipage}{0.90\textwidth}
\hspace{2.5em}
\centering
  \begin{minipage}{0.30\textwidth}
       % lelf lower right up trim= 7.5mm 0mm 0mm 10mm
   \includegraphics[height=3.0cm, width=\linewidth]{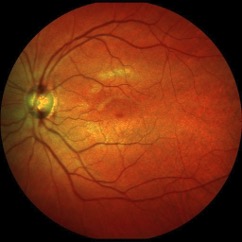}
   \centering \footnotesize \textbf{IMAGE}
  \end{minipage}
\begin{minipage}{0.30\textwidth}
\centering
       % lelf lower right up
  \includegraphics[height=3.0cm, width=\linewidth]{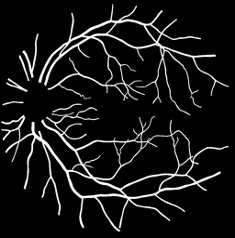}
   \footnotesize \textbf{GROUND TRUTH}
  \end{minipage}
  \begin{minipage}{0.30\textwidth}
  \centering
       % lelf lower right up
  \includegraphics[height=3.0cm, width=\linewidth]{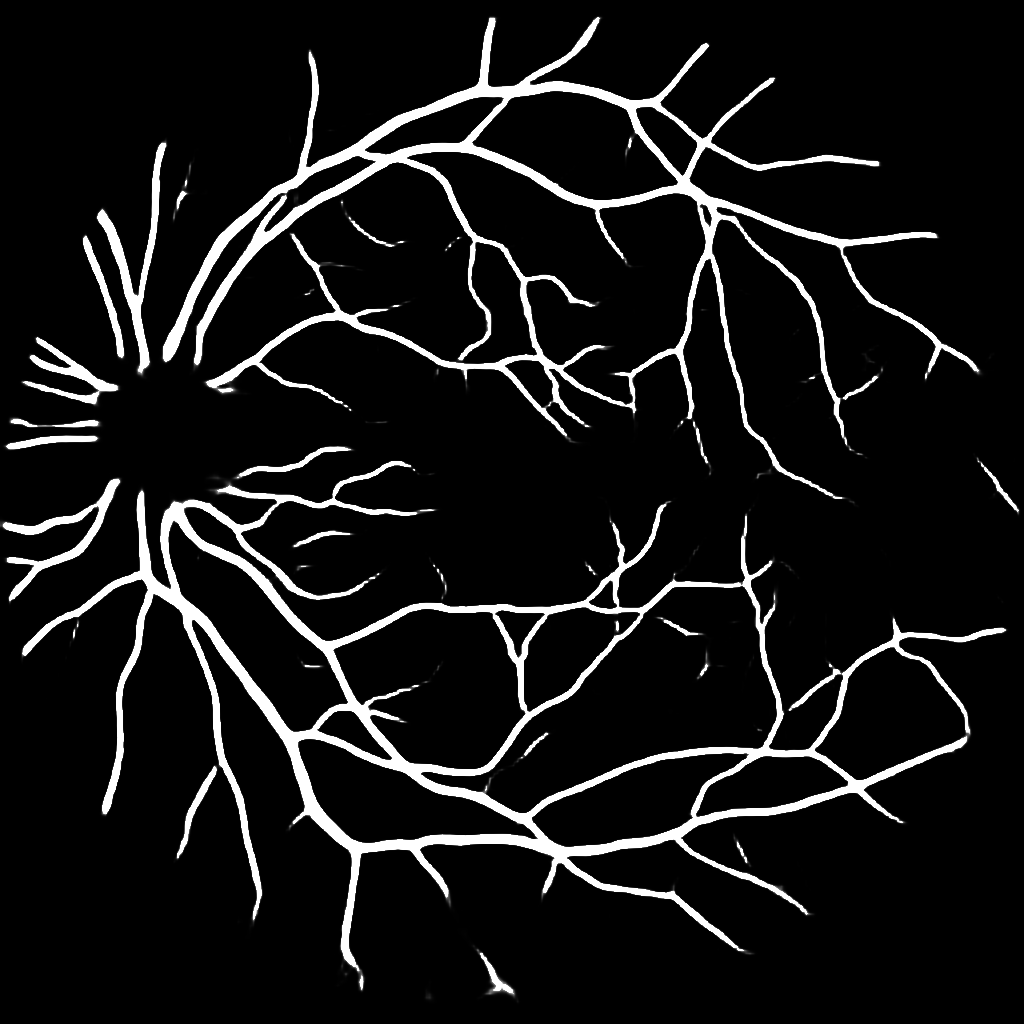}
  \footnotesize \textbf{PREDICTED MASK}
  \end{minipage}
 \vfill
  \begin{minipage}{0.96\textwidth}
  \centering
\caption{ \footnotesize{Visualizations of retinal vessel segmentation}}\label{fig:retinal-results}
\vspace{-1em}
  \end{minipage}
  \end{minipage}
 \end{figure*}

\begin{figure*}[!ht]
 \begin{minipage}{0.90\textwidth}
\hspace{2.5em}
\centering
  \begin{minipage}{0.30\textwidth}
       % lelf lower right up trim= 7.5mm 0mm 0mm 10mm
   \includegraphics[height=3.0cm, width=\linewidth, angle=180]{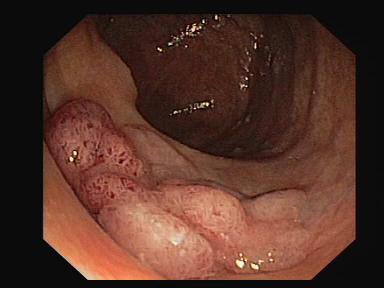}
  \end{minipage}
\begin{minipage}{0.30\textwidth}
      \centering
       % lelf lower right up
  \includegraphics[height=3.0cm, width=\linewidth, angle=180]{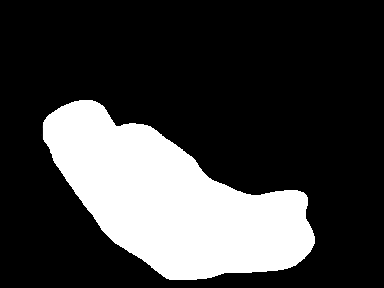}
  \end{minipage}
  \begin{minipage}{0.30\textwidth}
      \centering
       % lelf lower right up
  \includegraphics[height=3.0cm, width=\linewidth, angle=180]{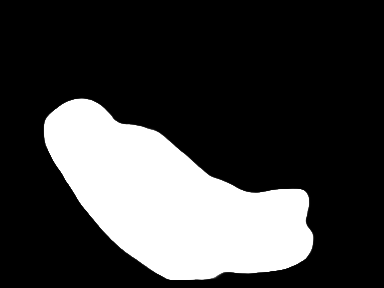}
  \end{minipage}
  \end{minipage}
\vfill
\begin{minipage}{0.90\textwidth}
\hspace{2.5em}
\centering
  \begin{minipage}{0.30\textwidth}
       % lelf lower right up trim= 7.5mm 0mm 0mm 10mm
   \includegraphics[height=3.0cm, width=\linewidth]{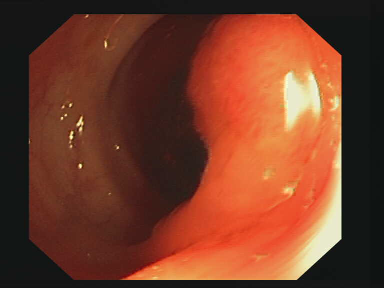}
  \end{minipage}
\begin{minipage}{0.30\textwidth}
\centering
       % lelf lower right up
  \includegraphics[height=3.0cm, width=\linewidth]{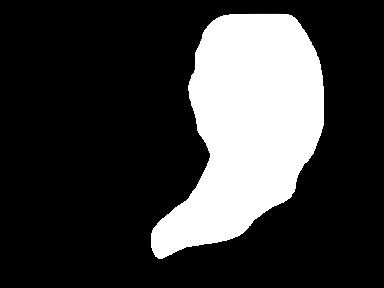}
  \end{minipage}
  \begin{minipage}{0.30\textwidth}
  \centering
       % lelf lower right up
  \includegraphics[height=3.0cm, width=\linewidth]{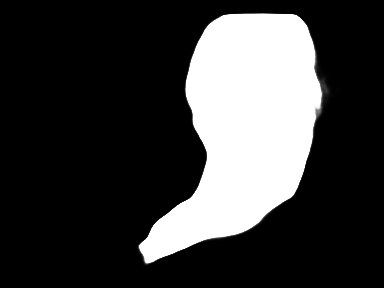}
  \end{minipage}
  \end{minipage}
  \vfill
  \begin{minipage}{0.90\textwidth}
\hspace{2.5em}
\centering
  \begin{minipage}{0.30\textwidth}
       % lelf lower right up trim= 7.5mm 0mm 0mm 10mm
   \includegraphics[height=3.0cm, width=\linewidth]{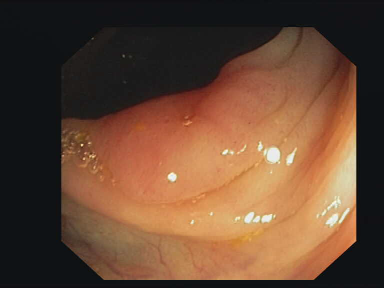}
  \end{minipage}
\begin{minipage}{0.30\textwidth}
\centering
       % lelf lower right up
  \includegraphics[height=3.0cm, width=\linewidth]{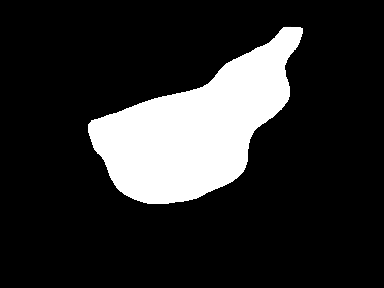}
  \end{minipage}
  \begin{minipage}{0.30\textwidth}
  \centering
       % lelf lower right up
  \includegraphics[height=3.0cm, width=\linewidth]{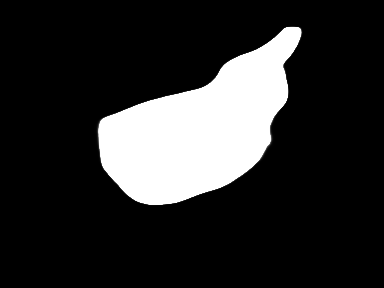}
  \end{minipage}
  \end{minipage}

  \vfill
  \begin{minipage}{0.90\textwidth}
\hspace{2.5em}
\centering
  \begin{minipage}{0.30\textwidth}
       % lelf lower right up trim= 7.5mm 0mm 0mm 10mm
   \includegraphics[height=3.0cm, width=\linewidth]{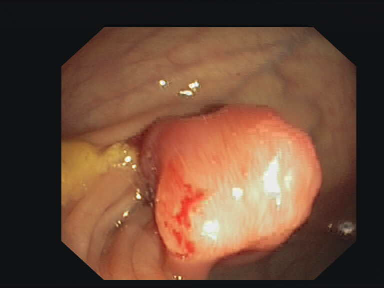}
   \centering \footnotesize \textbf{IMAGE}
  \end{minipage}
\begin{minipage}{0.30\textwidth}
\centering
       % lelf lower right up
  \includegraphics[height=3.0cm, width=\linewidth]{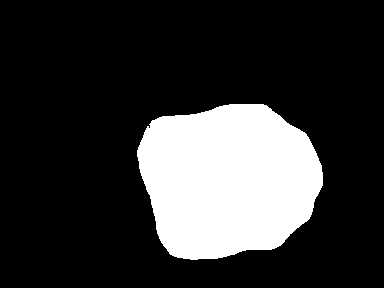}
   \footnotesize \textbf{GROUND TRUTH}
  \end{minipage}
  \begin{minipage}{0.30\textwidth}
  \centering
       % lelf lower right up
  \includegraphics[height=3.0cm, width=\linewidth]{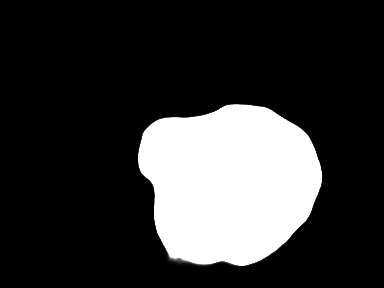}
  \footnotesize \textbf{PREDICTED MASK}

  \end{minipage}
  
      \vfill
  \begin{minipage}{0.96\textwidth}
  \centering
\caption{ \footnotesize{Visualizations of polyp segmentation}}\label{fig:polyp-results}
  \end{minipage}
  \end{minipage}
 \end{figure*}

We have provided the additional qualitative results of our method for all three segmentation tasks. Fig. \ref{fig:skin-results}, shows the qualitative results of skin lesion segmentation, while Fig. \ref{fig:retinal-results} and Fig. \ref{fig:polyp-results}, illustrate the visualization of retinal vessel and polyp segmentation tasks respectively. 

\end{document}